\documentclass[9pt, conference]{IEEEtran}
\IEEEoverridecommandlockouts
\usepackage{fancyhdr}
\usepackage[normalem]{ulem}
\usepackage{url}
\usepackage{tikz}
\usepackage{amsthm}
\usepackage{amsmath,amsfonts}
\usepackage{graphicx}
\usepackage{textcomp}
\usepackage[table,xcdraw]{}
\usepackage{booktabs} % For formal tables
\usepackage{stfloats}
\usepackage{subfigure}
\usepackage{multirow}
\usepackage{paralist}
\usepackage{balance}
\usepackage{bm}
\usepackage{pifont}
\usepackage{subcaption}
\usepackage{url}
\usepackage[hidelinks]{hyperref}

\usepackage[bottom=2.1cm, top=1.20cm, left=1.72cm, right=1.72cm]{geometry}
\usepackage{xspace}
\usepackage{algorithmic}
\usepackage[ruled,linesnumbered]{algorithm2e}
\usepackage{diagbox}
\usepackage{threeparttable}
\usepackage{makecell}
\usepackage{colortbl}
\usepackage{bm}
\definecolor{grey}{RGB}{0.5,0.5,0.5}
\usepackage{tcolorbox}
\newcommand{\parlabel}[1]{{\noindent\bf #1}}

\newcommand{\ie}{i.e.\xspace}

\newrobustcmd*\circled[1]{\tikz[baseline=(char.base)]{
            \node[shape=circle,draw,inner sep=1pt,fill,text=white,minimum size=1em] (char) {\textsf{\small #1}};}}

\usepackage{amsmath,amssymb,amsfonts}

\usepackage{cite}
\usepackage{caption}
\usepackage{colortbl}
\usepackage{tabularx}
\usepackage{textcomp}
\definecolor{darkgreen}{rgb}{0, 0.6, 0}

\begin{document}

% {RGB}{66, 186, 71}

\newcommand{\approach}{UVLLM}
\newcommand\name{UVLLM}

\title{UVLLM: An Automated Universal RTL Verification Framework using LLMs\vspace{-5pt}
\author{
    \IEEEauthorblockN{
        Yuchen Hu$^{1,2}$, Junhao Ye$^{1,2}$, Ke Xu$^{1,2}$, Jialin Sun$^{1,2}$, Shiyue Zhang$^{1,2}$, Xinyao Jiao$^{1,2}$, Dingrong Pan$^1$, Jie Zhou$^{1,2}$, \\Ning Wang$^3$, Weiwei Shan$^{1,2}$, Xinwei Fang$^4$, Xi Wang$^{1,2}$, Nan Guan$^3$, Zhe Jiang$^{1,2}$
    }
    \IEEEauthorblockA{
        $^1$National Center of Technology Innovation for EDA, China $^2$School of Integrated Circuits, Southeast University, China\\
        $^3$Department of Computer Science, City University of Hong Kong, Hong Kong\\
        $^4$Department of Computer Science, University of York, UK
    }
}
% \\
}

% \and
% \and
% \and
% \and
% \and

\maketitle

\begin{abstract}
% Hardware design verification is a critical yet labor-intensive and time-consuming process in the development of digital hardware systems. To alleviate these challenges, this paper introduces a novel framework, \textbf{\name}, which integrates Large Language Models (LLMs) with the established Universal Verification Methodology (UVM).
% By synergizing scripts with LLMs, our approach significantly enhances the automated testing and repairing for error-prone Register Transfer Level (RTL) codes. Specifically, all errors are triggered during verification, and \approach\  achieves a syntax error fix rate of 86.99\% and a functional error fix rate of 71.92\% on our proposed error benchmark, thus markedly improving the efficiency of the verification process compared with existing methods.
% Furthermore, this study also highlights the current limitations of LLM applications, particularly their dependence on extensive training data. 
% Our findings underscore the transformative potential of LLMs in hardware design verification, pointing to a promising direction for future research in AI-driven hardware design methodologies.

Verifying hardware designs in embedded systems is crucial but often labor-intensive and time-consuming. While existing solutions have improved automation, they frequently rely on unrealistic assumptions. To address these challenges, we introduce a novel framework, \name\, which combines Large Language Models (LLMs) with the Universal Verification Methodology (UVM) to relax these assumptions. \name\ significantly enhances the automation of testing and repairing error-prone Register Transfer Level (RTL) codes, a critical aspect of verification development. Unlike existing methods, \name\ ensures that all errors are triggered during verification, achieving a syntax error fix rate of 86.99\% and a functional error fix rate of 71.92\% on our proposed benchmark. These results demonstrate a substantial improvement in verification efficiency. Additionally, our study highlights the current limitations of LLM applications, particularly their reliance on extensive training data. We emphasize the transformative potential of LLMs in hardware design verification and suggest promising directions for future research in AI-driven hardware design methodologies.
The Repo. of dataset and code:
\textbf{\url{https://anonymous.4open.science/r/UVLLM/}}.
\end{abstract}

%%%%%%%%%%%%%%%%%%%%%%%%%%%%%%%
% In the semiconductor industry, Large Language Models (LLMs) have proven to be error-prone in generating Register Transfer Level (RTL) programming.
% In coping with this issue, researchers are investigating the potential of LLM-based debugging tools to identify and fix the bugs in the RTL code.
% However, due to the inherent uncertainties of LLMs and the intrinsic complexities of hardware design, existing solutions are still inadequate.

% \begin{IEEEkeywords}
% LLM, Hardware Design Verification, APR, UVM
% \end{IEEEkeywords}

\setlength{\textfloatsep}{9pt}% Remove \textfloatsep
\vspace{-2.5pt}
\section{Introduction}
\label{sc:Intro}
\vspace{-2.5pt}
% \zhenote{[British English or American English?]\\} -- 
% \zhenote{Check the experimental results across all the paper \\}

\begin{figure}[t]
    \centering
    \hspace{5pt}

    \subfigure[Workflow of the Frontend Hardware Design.]{\includegraphics[width=1\columnwidth]{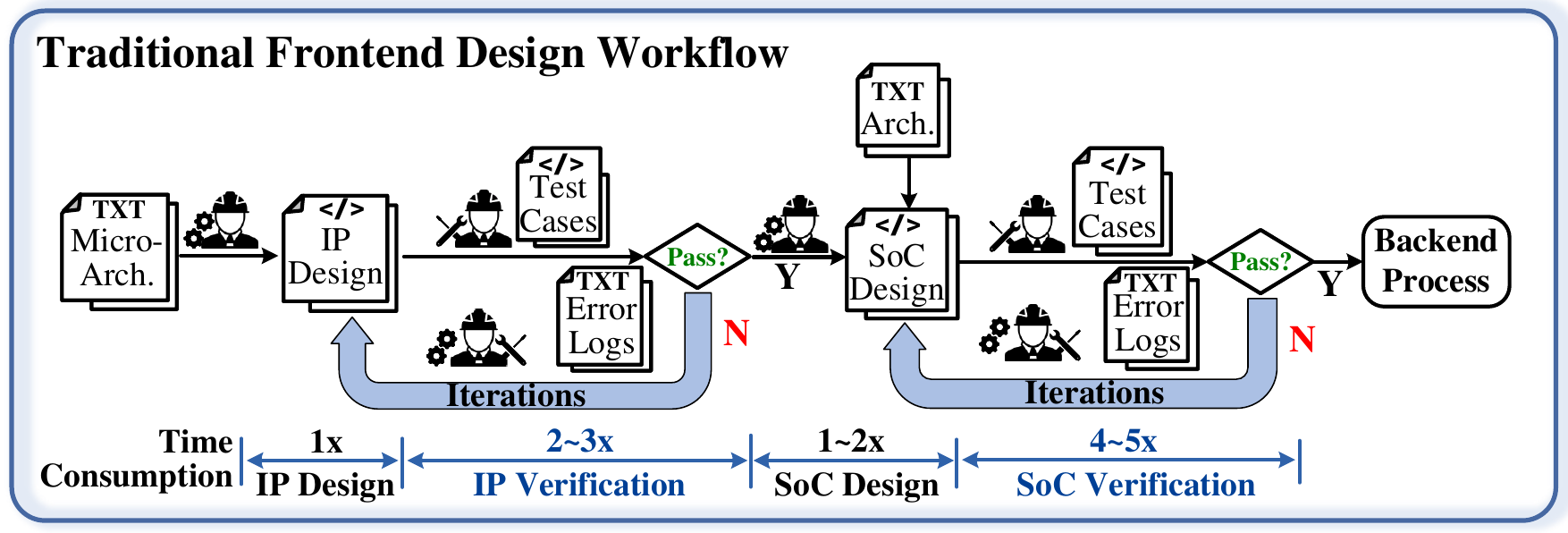}
    \label{fig:wf_fe}}
    \subfigure[Workflow of the APR System.]{\includegraphics[width=1\columnwidth]{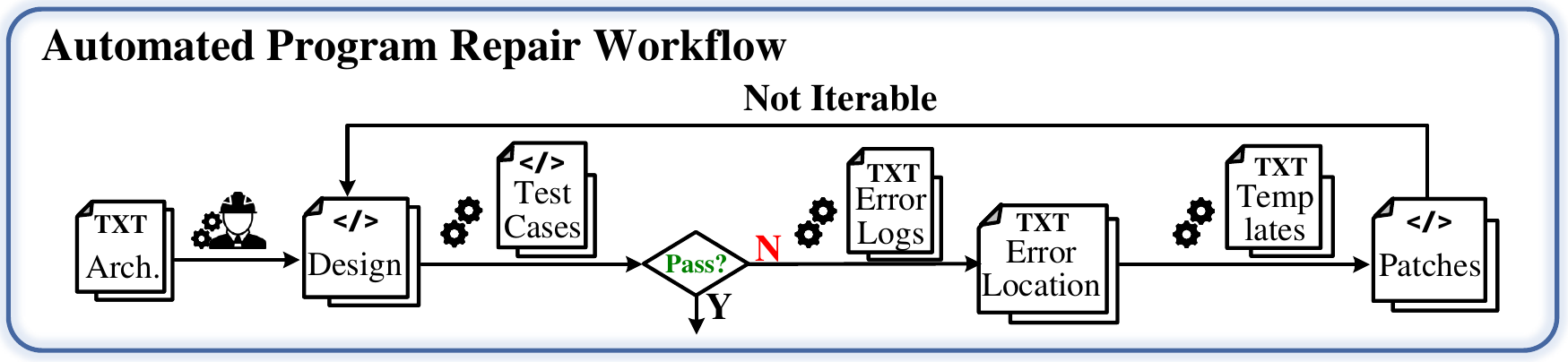}
    \label{fig:wf_apr}}
    % \caption{Workflow of the Frontend Hardware Design. The process is divided into initial design and verification phases, with verification accounting for more than 70\% of the frontend duration~\cite{lahti2018we}. Advanced APR techniques are integrated to automate and accelerate the repair stage during verification phase.
    %  ``$<$/$>$'' represents codes, as in the figures below.
    % }
    \vspace{-3pt}
    \caption{The frontend process is divided into initial design and verification phases, with verification accounting for more than 70\% of the  duration~\cite{lahti2018we}. Advanced APR techniques are integrated to automate and accelerate the repair stage during verification phase ($<$/$>$: RTL codes).
    }
    \label{fig:workflow}
    \vspace{-5pt}
\end{figure}

Hardware design verification in embedded systems remains heavily dependent on human expertise, making it a tedious and error-prone process that often incurs significant costs~\cite{lahti2018we}, as Fig.~\ref{fig:wf_fe} illustrates. 
% By reducing human involvement through effective verification strategies, there are substantial benefits such as minimising human errors and significantly reducing verification costs.
% %Typically, design verification unfolds in four key stages: Testbench Construction and Stimulus Generation;  Test Execution and Monitoring; Results Analysis; Debugging and Repair.
% The crucial part of this process is to debug and repair the design. Consequently, the burgeoning interest in automated program repair (APR) within software engineering, highlighted by recent studies~\cite{liu2018survey, liu2021critical,zhang2023critical, yin2024thinkrepair, xu2024automated}, is extending into hardware design. 
% The key component of this process involves debugging and repairing errors, areas where automated program repair (APR) can significantly contribute.
A critical aspect of this process is debugging and repairing errors, an area where automated program repair (APR) can contribute.
Originally developed for software~\cite{liu2018survey, liu2021critical,zhang2023critical, yin2024thinkrepair, xu2024automated}, APR uses automated tools to fix errors with minimum human intervention and is now being adapted for Hardware Description Language (HDL) design verification due to its potential to reduce human errors and verification costs. 

% Modern digital hardware design employs Hardware Description Languages (HDLs) such as Verilog and VHDL to delineate circuits at the Register Transfer Level (RTL), which are subsequently synthesized into physical circuits. 
APR systems, as depicted in Fig.~\ref{fig:wf_apr}, receive design codes and test cases, and attempt to enact targeted modifications with predefined templates to ensure all tests are passed.
Innovations such as Cirfix~\cite{ahmad2022cirfix}, Strider~\cite{yang2024strider}, and RTLrepair~\cite{laeufer2024rtl} demonstrate the potential of APR to reduce the labor and time required for hardware design verification. 
% These tools are integral to enhancing the verification process, ensuring that functional accuracy is maintained, which are critical for the subsequent phases of hardware synthesis and implementation.
However, these APR methodologies predominantly rely on fixed templates and focus on addressing functional error, limiting their scope and effectiveness of the repairs.
% , potentially compromising the verification quality
% Moreover, they focus on primarily addressing functional errors but often neglect syntax errors, which are equally important in harware verification. 
% Moreover, they are generally tailored to address functional defects and fail to correct syntax errors, which are crucial for ensuring the syntactic correctness required in verification.

% Recent advancements, particularly with the integration of Large Language Models (LLMs), have demonstrated promising results in automating key steps of the verification process. These include automated program repairs [cite], generating hardware specifications from code [cite], and debugging hardware designs for both functional and syntax errors [cite]. However, existing solutions along are insufficient for hardware verification due to their reliance on unrealistic assumptions and lack of consideration for the limitations inherent in LLMs. For example...

% Benefiting from the rapid development of LLMs, these challenges have found new solutions.
Fortunately, recent advancements in LLMs such as generating hardware code from natural language specifications~\cite{liu2023verilogeval,thakur2023verigen,blocklove2023chip,delorenzo2024make}, and debugging hardware designs for both functional and syntax errors~\cite{liu2023chipnemo, tsai2023rtlfixer, yao2024hdldebugger, xu2024meic, fu2023llm4sechw}, have demonstrated promising results in bridging this gap.
% automating key steps of the verification process. 
% These include  generating hardware code from natural language specifications~\cite{liu2023verilogeval,thakur2023verigen,blocklove2023chip,delorenzo2024make}, and debugging hardware designs for both functional and syntax errors~\cite{liu2023chipnemo, tsai2023rtlfixer, yao2024hdldebugger, xu2024meic, fu2023llm4sechw}. 
However, existing solutions are still insufficient for hardware design verification due to their reliance on unreliable assumptions and lack of consideration for the limitations inherent in LLMs. 
% It becomes a big issue about how to reveal the errors.
For instance, despite demonstrations of high fix rates for both syntax and functional errors~\cite{xu2024meic}, our analysis indicates that approximately 10\% of the benchmarks manage to bypass the testbench without undergoing any repairs, and the reliability of some repairs remains questionable owing to insufficient coverage of test cases. These findings underscore the need for more robust solutions capable of effective deployment in real-world verification scenarios.

% the MEIC framework~\cite{xu2024meic}, which employs modified LLM agents, demonstrates high fix rates for both syntax and functional errors. Yet, approximately 10\% of the benchmarks still escape the testbench without any repairs, and the reliability of some repairs is questionable due to insufficient test case coverage.  
% These issues highlight that current solutions may not be robust enough for deployment in real-world scenarios, where the complexity and variability of practical environments can expose the inadequacies of the existing testing protocols.

% To conclude, the prevailing research on LLM-aided debugging tends to overly highlight capabilities of LLMs but neglects the crucial need for systematic verification consisting of testing and repairing.
% Regarding this issue, we propose a comprehensive end-to-end hardware design verification framework, Universal Verification via Large Language Model (UVLLM),  that integrates the established UVM with emerging and off-the-shelf methods, algorithms, and tools. This approach enables the first automated hardware design verification framework that operates under practically realistic assumptions while uncertainties associated with LLM behaviour. 
To address these shortcomings, we propose a comprehensive end-to-end hardware design verification framework, Universal Verification via Large Language Model (UVLLM). This framework integrates the established UVM with LLMs. Our approach enables the first automated hardware design verification framework that operates with practically assumptions, while also managing uncertainties associated with LLM behaviors. This systematic verification strategy, which encompasses testing and repairing, ensures a more robust solution than those currently highlighted in LLM-aided debugging research.

% \parlabel{Contributions.}
%We present \textbf{Universal Verification via LLMs (\name)}, a novel automated-hardware-verification (\textbf{AHV}) framework that utilises multiple LLMs to enable automated and iterative verification of RTL code. \name\ is designed to address the above limitations commonly associated with applying LLMs in hardware verification. 
% We present \textbf{Universal Verification via LLMs (\name)}, a new Automated Hardware Verification (AHV) framework that combines LLMs with the UVM to enable automated and iterative verification of RTL code. \name\ is designed to overcome the common limitations associated with using LLMs in hardware verification.
The main \textbf{contributions} of this paper are:
\vspace{-3pt}
\begin{itemize}
    %\item \textbf{A reliable verification framework:} \name\ integrates full-stack verification methods, with multi LLM agents, and code repositories. This allows continuous verification and evaluation of RTL code, mitigating uncertainties caused by the fluctuations in the performance of LLM outputs.
    % \item \textbf{A reliable verification framework:} \name\ synergistically combines LLMs with the UVM, thereby creating a unified and scalable verification environment, enabling the automated and iterative verification of RTL code.

    \item \textbf{A comprehensive testing: }Utilizing the UVM, our approach enables flexible test modes and efficient coverage collection.  Additionally, the reference models generated by LLMs provide a robust foundation for testing across diverse input scenarios.    % \item \textbf{Dual LLM deployment:} \name\ employs a fine-tuned debug agent that identifies and attempts to correct syntax and function errors, followed by an LLM scoring agent that assesses the quality of RTL candidates derived from the debug agent. This deployment provides quantified and traceable feedback that informs further iterations.
    
    \item \textbf{An open-sourced tooling:} We have created an open-source toolset to enable the broad and early adoption of \name, thereby easing its integration. These tools are publicly available at \textbf{\url{https://anonymous.4open.science/r/UVLLM/}}.

    % \item \textbf{Joint LLM-Script Method:} \name\ incorporates a LLM-Script-mixed method. With the LLM-based, script-supplemented strategy, \name\ enables a robust, token-effective systematic verification methodology.
    % bug localization engine to post-process logs generated by the UVM testbench. This engine systematically extracts relevant data from the logs, providing essential insights that facilitate the debugging process and enable the precise identification of design issues.
    
    \item \textbf{Extensive empirical validation:} We present an open-source error dateset derived from verified projects, containing 331 code instances with realistic errors across various modules, generated by our paradigm error generator.  We will continue to organize and update this dataset periodically.
    
    \item \textbf{Demonstrated performance improvement:} \name, incorporating GPT-4-turbo, significantly enhances verification automation, achieving a syntax error fix rate of 86.99\% and a functional error fix rate of 71.92\%, exceeding MEIC~\cite{xu2024meic} in terms of repair rates and execution time under realistic testing scenarios. 
    % Also, i
    It delivers up to 48x speedup in debugging processes when compared with experienced engineers. 
\end{itemize}

\parlabel{Organisation.} The structure of this paper is as follows: Section~\ref{sc:Concepts} outlines the fundamental concepts of \name, while Section~\ref{sc:Details} delves into the details of \name\ and their underlying reasons. 
Section~\ref{sc:Evaluation} assesses our framework and compares it with existing methods. 
Section~\ref{sc:Conclusion} provides conclusions and make discussions.
% The rest of the paper is organised as follows: Section~\ref{sc:Concepts} presents the top-level concepts of \name, Section~\ref{sc:Details} introduces the details of \name\ and the rationales.
% Section~\ref{sc:Evaluation} evaluates our framework conducts a comparative analysis with existing methodologies. 
% %followed by the related work given in Section~\ref{sc:RelatedWork}. 
% Section~\ref{sc:Conclusion} concludes and offers the insights.

% \textcolor{red}{Finally, we conclude the paper with the lessons we learnt in Section~\ref{sc:Conclusion}.}

\vspace{-2.5pt}
\section{\name: An Overview}
\label{sc:Concepts}
\vspace{-2.5pt}

\begin{figure*}[t]
    \centering
    \includegraphics[width=1.0\textwidth]{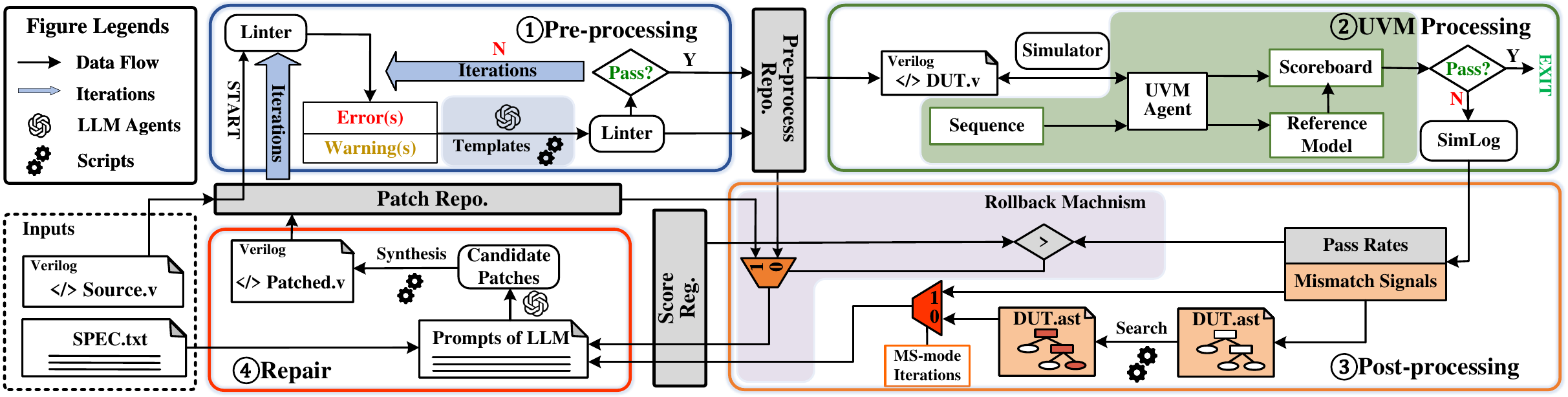}
    \caption{
    The \name\ Framework Overview: The process begins with the DUT and the Specification (Spec.), where the Spec. is used to generate a reference model. Initially, the DUT is pre-processed to eliminate syntax and focused timing-related errors (Step \textcircled{\scriptsize{1}}). Subsequently, the pre-processed code is tested under a UVM testbench (Step \textcircled{\scriptsize{2}}), and the log is then post-processed to extract relevant data (Step \textcircled{\scriptsize{3}}), which the debug agents use to generate candidate patches (Step \textcircled{\scriptsize{4}}). These codes and their pass rates are archived in the Repository (Repo.) and Register (Reg.) for future iterations.    
    }
    \hspace{-15pt}
    \vspace{-15pt}
    \label{fig:overview}
\end{figure*}

Designed to enhance the hardware design verification phase, \name\ aids hardware developers in detecting and correcting common errors in RTL code. The \name\ is applicable to various hardware environments, in this work, we illustrate the usage of \name\ in Verilog. As depicted in Fig.~\ref{fig:overview}, this framework combines traditional UVM with LLMs and assumes following inputs:

\begin{itemize}
%\vspace{-1.5pt}
    \item \textbf{Design specifications} that outline the intended and expected behavior of the hardware component;
    
    \item \textbf{RTL codes} that contain the untested RTL code of the initial hardware design, \ie, Design Under Test (DUT).
        
\end{itemize}
%\vspace{-1.5pt}

Reflecting advancements in state-of-the-art (SOTA) research, it is noted that using LLMs for hardware design verification requires an iterative approach~\cite{xu2024meic} and is more effective when LLMs are provided with detailed error information~\cite{white2023prompt,sahoo2024systematic,wei2022chain}. 
However, using LLMs still presents certain shortcomings, as current applications of LLMs struggle with reliability, applicability, and processing long code sequences.
Additionally, the use of SOTA LLMs can be costly(e.g., the GPT-4-Turbo model, charges \$0.01 per $1K$ input tokens and \$0.03 per $1K$ output tokens~\cite{openai2024gptapi}).
%Despite these advancements, challenges remain: current LLMs struggle with reliability, applicability, and handling long code sequences. Moreover, their use can be prohibitively expensive; for instance, the GPT-4-Turbo model charges \$0.01 per $1K$ input tokens and \$0.03 per $1K$ output tokens~\cite{openai2024gptapi}).
%To address these limitations, \approach\ aims to verify RTL code against design specifications through a structured four-step process, as illustrated in Fig.~\ref{fig:overview}. 
To address these challenges, \approach\ introduces a cost-efficient, structured four-step process for verifying RTL code against design specifications, as illustrated in Fig.~\ref{fig:overview}.
% \approach\ hypothesizes that the LLMs can perform better based on more detailed error information. Our approach for the processing of error information is discussed in Section \ref{sc:Details}.
% Moreover, previous work has proven that debugging with LLMs is not an one-shot process. Under these preliminaries, \approach\ attempts to correct the RTL code as necessary across iterations from four pipeline stages (Step \textcircled{1}-- \textcircled{4}) in Fig.~\ref{fig:overview}. 
% The iterative \approach\ pipeline involves the following steps:

% \begin{itemize}

    \textit{1) \textbf{\quad Pre-processing: }} Starting with raw RTL code as the input, this stage utilizes linters such as Verilator to pre-process the code, removing syntax errors and addressing timing-related functional errors using a combined LLM-script method. 
    The output is  syntax-correct DUT, setting a solid foundation for further functionality testing.

    \textit{2) \textbf{\quad UVM Processing: }} Pre-processed DUT code is then tested against a pre-built UVM testbench to identify behavioral discrepancies. Outputs include detailed logs that either confirm alignment with the reference model or highlight deviations with specific signal values and test pass rates, facilitating targeted repairs.

    \textit{3) \textbf{\quad Post-processing: }} Utilizing the UVM logs as input, this stage analyzes the logs to extract critical error data using a localization engine and the Abstract Syntax Tree (AST). The output isolates mismatch signals and specific error paths, preparing them for precise correction in the repair stage.

    \textit{4) \textbf{\quad Repair: }} The final stage takes the design description, the DUT code and the detailed error information from the post-processing as input. Utilizing the information, the debug agents offer candidate patches to correct the errors.
    % , aligning the DUT code with the reference model’s behavior. 
    The repaired DUT code is then synthesized as the stage output for further iteration.

The termination conditions for the framework loop are: \textbf{1)} no errors are detected (\textbf{Success}), or \textbf{2)} the maximum number of iterations is reached (\textbf{Failure}). If any of the above conditions are met, the iteration stops. All history files are stored for reference.
% , and the most recent code is output.

\parlabel{Modularization.} \approach\ uses modular design which allows it to adapt to a wide range of verification scenarios by enabling the use of different tools based on the needs. For example, one can replace an LLM with a more advanced model or use a different linter if required. This flexibility is made possible by standard interfaces between the pipelines, which simplify the integration of different tools via adjustments to the API or by keeping a consistent format.

\vspace{-2.5pt}
\section{\name: The Framework Pipeline}
\label{sc:Details}
\vspace{-2.5pt}

As for the operation of the framework, we introduce the joint LLM-Script pre-processing stage (Section \ref{sbsc:prep}), processing stage with tests (Section \ref{sbsc:ref}), post-processing stage for error location (Section \ref{sbsc:loca}), followed by the discussion on the microsystems integrated with the LLM agents (Section \ref{sbsc:de}). At last, we present our effort for human-like error generation (Section \ref{sbsc:data}) for evaluation.

% Medialization --- Input, output, and why?

% \begin{itemize}
%     \item Fine-Tuning
%     \item Local RTL simulation with RTL source code and Golden Reference
%     \item Website
%     \item LLM debug
%     \item Iteration
%     \item Evaluation and convergence
% \end{itemize}

\begin{algorithm}[t]
{\small
\SetAlgoLined
% $\vartriangleright$ \text{\texttt{Pre-processor}}\\
\KwIn{DUT file $F_{D}$}
\KwOut{Pre-processed DUT file $F_{Dprep}$}
\SetKwFunction{FMain}{\normalfont PreproDUT}
    \SetKwProg{Fn}{Function}{:}{}
    \Fn{\FMain{\normalfont $F_D$}}{
    {
        % $F_{Dtemp}$ = $F_D$;\\
        % \While{$Errors\ \| \ Warnings$}{
        \Repeat{$(Errs == \varnothing) \& (Warns == \varnothing)$}{
            $Log$ = Linter($F_{Dprep}$);\\
            $Errs$ = Match($Log$, $Error$);\\
            $Warns$ = Match($Log$, $Warning$);\\
            \uIf{$Errs$}{
            $F_{Dprep}$ = GPT($F_{Dprep}$, $Errs$);\\} 
            \ElseIf{$Warns$}{
            $WarnTemps$ = Search($Warns$, $WarnList$);\\
            $F_{Dprep}$ = Replace($F_{Dprep}$, $WarnTemps$);\\
            }
        }
        % $F_{Dprep}$ = $F_{Dtemp}$;\\
        \textbf{return}\ $F_{Dprep}$\\
        % MT, MS = getMismatch($L_{UVM}$, $PAT_{MS}$);\\
        % \If{MS}{
        %     IV = getInputValue($W_D$, MT);\\
        %     \textbf{return}\ MT, MS, IV\\
        % }
        % {\textbf{return}\ 0, $\varnothing$, $\varnothing$\\}
} }
\textbf{End Function}   
}
\caption{Pre-processing DUT with Joint LLM-Script.}
\label{algo:prep}
\end{algorithm}

\vspace{-2.5pt}
\subsection{Pre-processing via the Linter}
\label{sbsc:prep}
\vspace{-2.5pt}
% Static localization in APR employs static code analysis and slicing to pinpoint potential error sources without executing the code. 
% LLMs have been demonstrated to be effective in Verilog debugging, especially for syntax error repair.
To ensure compliance with best practices and avoid obvious functional defects, the code is pre-processed using Verilator for linting before UVM testbench evaluation. 
%To improve code quality and ensure compliance with best practices, the code undergoes static analysis with Verilator before UVM testbench evaluation.
Through static code analysis and slicing, the method  pinpoints potential error sources without executing the code.
LLMs have proven effective in Verilog debugging, especially in repairing syntax errors and refining code with offered error sources.
% This step is crucial for catching syntax and potential semantic issues early, thus reducing the need for more resource-intensive verification process and competitive LLM use.
By resolving syntax errors and identifying semantic issues early on, the subsequent need for employing LLMs for debugging is minimized, reducing the  costs for the use of LLMs. 

\parlabel{Combined LLM-Script Pre-pocessing.}

To further reduce costs, as detailed in Algorithm \ref{algo:prep}, we employ a cost-effective strategy that combines LLMs with supplementary scripting to minimize unnecessary LLM usage. In the pre-processing stage, LLMs are only utilized to help identify and correct syntax errors. These models draw on their extensive training across diverse codebases, efficiently recognizing and amending a broad spectrum of common coding errors and integrating essential error information.

Additionally, the pipeline incorporates scripts designed to address specific warnings that, while not syntax errors, could lead to potential runtime issues, especially some timing-related ones. For instance, in combinational logic circuits, blocking assignments are typical, and  Verilator issues warnings for using non-blocking assignments in this case. Through predefined templates, it is possible to systematically identify and modify issues where a non-blocking assignment ``$<=$'' should be replaced with a blocking assignment ``$=$'', ensuring the code adheres to expected timing behaviors.
% optimizing the code for enhanced performance and reliability.

The pre-processing stage iterates until all syntax errors and focused timing-related warnings are resolved. Integrating  LLM agents with scripts establishes a robust foundation for subsequent verification stages, ensuring that the DUT encounters only functional errors.

\begin{figure}[t]
    \centering
    % \vspace{-7pt}
    \includegraphics[width=1.00\columnwidth]{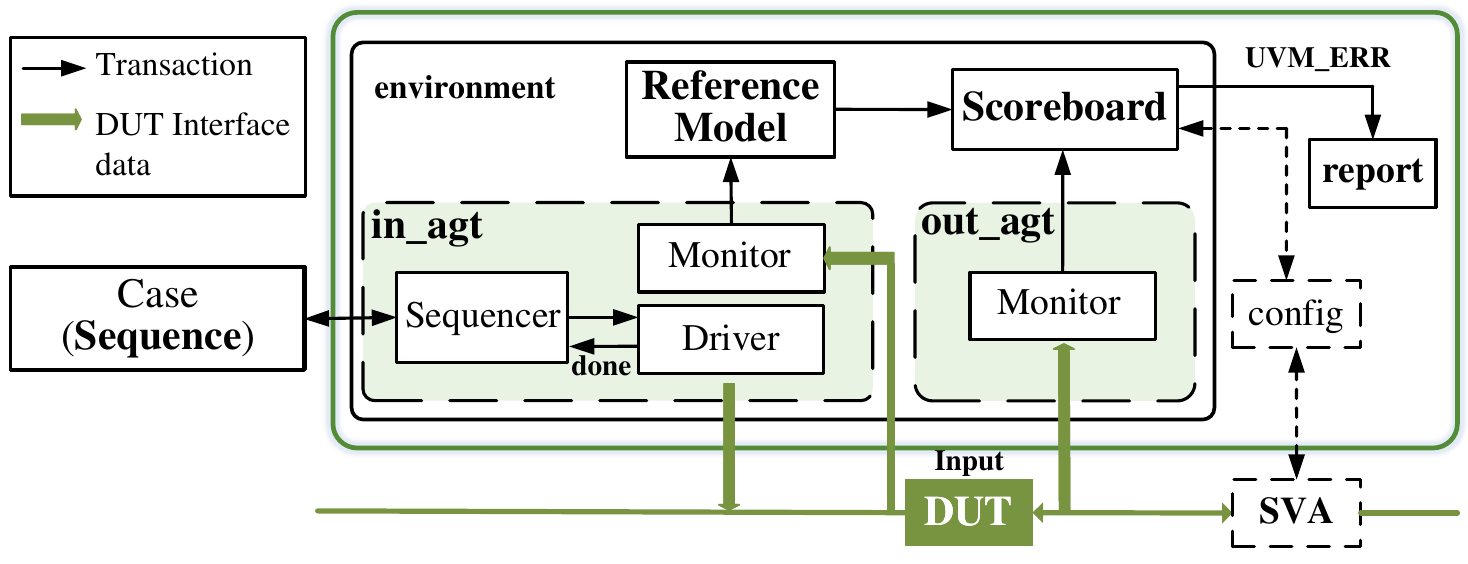}
    \caption{Structure of the UVM~\cite{plasencia2018robust}, illustrating the key components and their interactions within a typical verification environment.}
    \vspace{-5pt}
    \label{fig:uvm}
\end{figure}

\vspace{-2.5pt}
\subsection{UVM Processing}
\label{sbsc:ref}
\vspace{-2.5pt}
% \textbf{Claim the shortcoming of current work: 1.It is hard to develop testbench with human-written outputs for complex design (correctness validation, coverage); 2.Current work takes no universal form of testbench for massive production}

% \begin{figure}[t]
%     \centering
%     % \vspace{-7pt}
%     \includegraphics[width=1.00\columnwidth]{graphics/UVM_pipeline.pdf}
%     \caption{Structure of the UVM~\cite{plasencia2018robust}, illustrating the key components and their interactions within a typical verification environment}
%     % \vspace{-10pt}
%     \label{fig:uvm}
% \end{figure}

Recent studies, such as MEIC~\cite{xu2024meic}, validate LLMs' effectiveness in hardware debugging.
% However, these studies do not systematically assess the rationale for introducing LLMs at each stage of the debugging process, which raises questions about the appropriateness and potential inefficiencies of applying LLMs across various debugging tasks. 
However, these studies neglect the importance of testbench construction and employ finite test cases, which restricts the test coverage and leads to overfitting on specific cases. This limited scope significantly undermines the general applicability of the results, as evidenced by a \textbf{10\%} reduction in the actual fix rate reported by MEIC due to many error instances escaping detection.
To overcome these limitations, the UVM framework, depicted in Fig.~\ref{fig:uvm}, is employed as the testbench for \approach\ to verify RTL codes, due to its robust support for flexible and complete testing modes.
% Based on SystemVerilog, UVM is designed to support the creation of scalable, reusable, and robust test environments. 

\parlabel{UVM Construct.}
To verify complex hardware system, the UVM offers a formal verification structure significantly advanced over simpler testbenches, incorporating components like Agents, Environments, Sequencers, Drivers, and Monitors. Each agent encapsulates a sequencer, driver, and monitor, enabling direct interaction with the DUT. The Sequencer organizes transactions generated from Sequence that simulate real-world operations, which the Driver then translates into pin-level actions on the DUT.
Central to UVM's effectiveness is the Scoreboard, which compares actual results with expected outcomes to ensure the DUT performs correctly under various conditions. 
Additionally, UVM supports various test modes and coverage collection techniques that further enhance testing thoroughness. 
This method helps identify discrepancies and potential failures, enhancing the reliability and accuracy of the verification process.

\parlabel{Reference Model Generation.}
In UVM, reference models play a crucial role in verifying complex designs, such as those used in digital signal processing and cryptography, by providing high-level abstractions of the DUT. These models enhance simulation accuracy and efficiency, contributing to a more streamlined verification process. Traditionally, C/C++ is preferred for reference models in industry due to its seamless integration with SystemVerilog via Direct Programming Interfaces (DPI), which accelerates verification cycles~\cite{spear2008systemverilog,semeria2000methodology,nakamura2004fast}.  
In this context, the capabilities of LLMs are especially relevant. Given the abundance of open-source datasets, LLMs have shown remarkable proficiency in generating C/C++ code, making them well-suited to assist in crafting adaptable, high-quality reference models. These LLM-generated models can dynamically respond to the intricate demands of verification, continuously updating to support high-fidelity simulations and robust design validation.

\parlabel{Extensibility.}
The extensibility of the UVM is particularly evident when considering the integration of automated assertion generation. UVM's structured, modular framework for verification is optimally configured to incorporate advanced enhancements such as AI-driven assertions, which can systematically verify that the design behaves as expected across various protocols like APB (Advanced Peripheral Bus) and AHB (Advanced High-Performance Bus)~\cite{radu2024generative}.

\vspace{-3.5pt}
\subsection{Post-processing via Localization Engine}
\label{sbsc:loca}
\vspace{-2.5pt}

\begin{algorithm}[t]
{\small
\SetAlgoLined
% $\vartriangleright$ \text{\texttt{Localization Engine}}\\
\KwIn{DUT file $F_{D}$, UVM log $L_{UVM}$, reference waveform $W_R$, simulation waveform $W_S$, iterations $Iter$}
\KwOut{Error information $ErrInfo$}
\SetKwFunction{FMain}{\normalfont ErrChk}
    \SetKwProg{Fn}{Function}{:}{}
    \Fn{\FMain{\normalfont $L_{UVM}$, $W_S$}}{
    {
        % \textcolor{blue}{\textbf{\name}.b.check(DISABLE);}\\
        % \textbf{Kernel}.Intr(DISABLE);\\
        % task *\textit{next} = NULL;\\
        \emph{$/*\ MT:\ Mismatch\ Timestamp\ */$}\\
        \emph{$/*\ MS:\ Mismatch\ Signals\ */$}\\
        \emph{$/*\ IV:\ Input\ Values\ */$}\\
        $MT$, $MS$ = getMismatch($L_{UVM}$, $PAT_{MS}$);\\
        \If{MS}{
            
            $IV$ = getInputValue($W_S$, $MT$);\\           
        }
         \textbf{return}\ $MT$, $MS$, $IV$\\
        % {\textbf{return}\ 0, $\varnothing$, $\varnothing$\\}
} }
\textbf{End Function}   

\SetKwFunction{FMain}{\normalfont ErrInfoFetch}
    \SetKwProg{Fn}{Function}{:}{}
    \Fn{\FMain{\normalfont $F_D$, $L_{UVM}$, $W_R$, $W_S$, $Iter$}}{
    {       
        \emph{$/*\ SL:\ Suspicious\ Code\ Lines\ */$}\\
        $MT$, $MS$, $IV$ = ErrChk($L_{UVM}$, $W_S$);\\
        \For{$ms$ $\in$ $MS$}
        {
            $DFG$ = getDFG($F_D$, $ms$);\\
            
            $SL$ = $SL$ $\bigcup$ traverse($DFG$, $IV$);  \\
            % \tcc{Suspicious Code Lines}\\
            \If{detectSignal($FL$, $s$) and $s$ $\not \in$ $MS$}{
            $MS$ = $MS$ $\bigcup$ \{$s$\};\\
            }
        }
        $ErrInfo$ = ($Iter<TH$) ? $MS$ : $SL$;\\
        \textbf{return} $ErrInfo$\\ 
        % \eIf{$Iter < TH$}{
        % % $Ite += 1$;\\ 
        % \textbf{return}\ $MS$
        % \\}
        % {\textbf{return}\ $SL$\\}
    }
}
\textbf{End Function}
}
\caption{Post-processing with Localization Engine.}
\label{algo:postp}
\end{algorithm}

Current LLM-aided verification methods tend to use minimally processed logs as inputs, which are often low in information density, thereby diminishing the efficiency of LLMs in diagnosing and fixing errors. To address this, we adopted time-aware dynamic error localization~\cite{yang2024strider} to extract more concrete and high-value information from these logs with methods. 
% Unlike the static localization method described in Section~\ref{sbsc:prep}, which generally lacks precision and temporal sensitivity, the time-aware dynamic localization method offers a more refined approach. 
This method, tailored for HDL environments, surpasses the static localization method described in Section~\ref{sbsc:prep} by providing greater precision and temporal sensitivity. 

The localization engine leverages dynamic analysis and temporal insights to detect discrepancies between expected and actual signal outputs as recorded in the UVM log. These discrepancies are crucial for performing dynamic slicing through data flow graphs (DFGs), as outlined in Algorithm~\ref{algo:postp}.
 To optimize token usage, \approach\ adopts a segmented information extraction strategy. Initially, mismatch signals are input into the LLM's prompt as indicators of potential errors. If subsequent repair attempts fail, this indicates that relying solely on mismatched signals may be insufficient. To increase diagnostic precision, the system then incorporates actual execution paths—identified as suspicious—into the analysis alongside the error signals.
% To optimize token usage, a segmented information extraction strategy is adopted. Initially, mismatched signals are fed into the LLM's prompt as error indicators. However, it may become clear that wrong signals are insufficient if repair fails specific iterations. To improve accuracy, the actual execution paths, deemed suspicious, are included alongside the error signals through data flow graphs (DFGs). 
This approach focuses on actual execution paths in operation, leading to more precise identification of errors.

\parlabel{Rollback Mechanism.}
During the development of \name, a \emph{Rollback mechanism} was implemented to address the issue of inaccuracies in LLM outputs, often termed ``hallucination''~\cite{ji2023towards,amatriain2024measuring,galitsky2023truth,chen2023hallucination,ji2023survey}. This feature is crucial for preventing the accumulation of errors across iterations due to reliance on flawed candidate repairs. Despite previous studies utilizing LLMs as reward models to assess repair quality\cite{xu2024meic,lambert2024rewardbench}, there is a lack of robust quantitative metrics to effectively measure the correctness of these modifications. This gap can lead to inefficiencies in the rollback process, potentially triggering false positive rollbacks.

Within the UVM framework, the quality of Verilog code iteration is evaluated using a scoreboard that assigns the test pass rate. We assume that higher scores correlate with fewer errors and better functionality. 
% This score is indicative of the code's coverage and quality.
The Rollback mechanism functions by preserving a history of code versions and their scores. If a new iteration scores lower than a previous one, indicating a decline in code quality, the mechanism reverts to the highest-scoring version, as illustrated in Fig.~\ref{fig:overview}. The alterations that led to the decrement in score are thus recorded as ``damage repairs'', which is utilized in Fig.~\ref{fig:cofl}.
% This strategy not only prevents the accumulation of errors but also enhances the traceability and monitoring of code evolution.

\vspace{-2.5pt}
\subsection{Repair Agent}
\label{sbsc:de}
\vspace{-2.5pt}
\begin{figure}[t]
    \centering
    % \vspace{-5pt}
    \includegraphics[width=1.0\columnwidth]{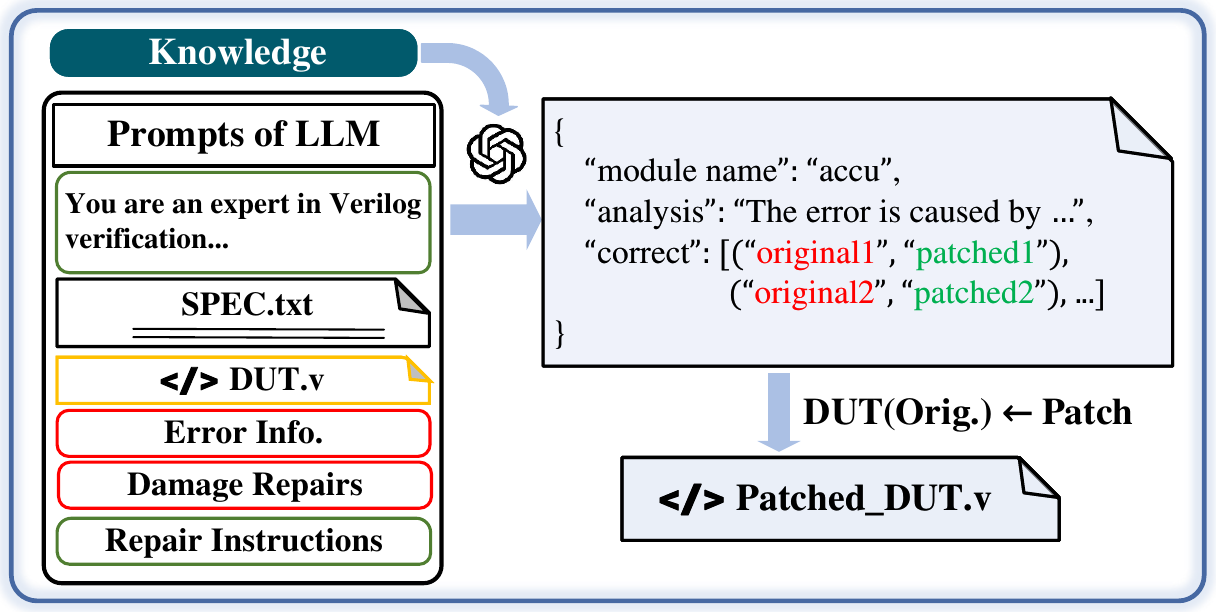}
    % \vspace{-10pt}
    \caption{Input and Output Formats for LLM Agents: The input format is structured for various agents with minor prompt modifications. The output is provided in JSON format, featuring original-patch pairs.}

    % \vspace{-5pt}
    \label{fig:cofl}
\end{figure}
The LLM agent functions as an adept RTL repair expert, leveraging three key inputs: the design specification, which outlines intended functionality and port definitions; RTL code; and error information. To enhance the repair process and facilitate iterative improvements, the system incorporates ``damage repairs'' as an additional input, crucial for preventing the recurrence of unsuccessful corrections when the rollback mechanism is activated. 
In a multi-agent setup, specific modifications to the prompts for each agent address different aspects of the debugging process, enabling a more nuanced and thorough error analysis in the RTL code. The primary prompts that guide the debugging activities of these agents are illustrated in Fig.~\ref{fig:cofl}, emphasizing a tailored approach to error resolution.

\parlabel{Formalizing agent's outputs.}
 It's commonly noticed that LLM-generated responses tend to include detailed explanations during debugging, which can clutter the main objective of code debugging. 
 In light of our observations, it becomes evident that LLMs exhibit enhanced debugging capabilities with superior reasoning process.
 % Figure~\ref{fig:cofl}. 
 To prevent the inclusion of irrelevant details and erase the hallucination during the iterative cycle, a method for distilling the responses generated by the LLM is adopted. Utilizing the Structured Outputs method enables adherence to JSON Schema~\cite{Structured-outputs}, thereby ensuring that responses are consistently formatted according to predefined structures. By requiring the response to be in JSON format and to contain an element labeled ``correct'' which consists of pair of wrong codes and right codes, the code sections accentuated in Fig.~\ref{fig:cofl} are refined and carried into the subsequent iteration.

\vspace{-1.5pt}
\subsection{Benchmark Generation}
\label{sbsc:data}
\vspace{-1.5pt}

The incidence of errors in module code is closely related to specific attributes such as code length and functional complexity~\cite{sudakrishnan2008understanding,ma2022debugging,antinyan2017evaluating}. To evaluate the efficacy of the verification methodology, a well constructed evaluation dataset was developed through a systematic selection of samples from validated open-source datasets, representing a diverse range of codebases. These samples were then deliberately infused with typical errors to simulate real-world coding mistakes.

In design bases combining both commercial and open-source IPs, a comparative analysis was performed on two consecutive versions of the code: one immediately before and another after code repository commits. This analysis focused on identifying discrepancies and documenting changes made during the design process, effectively highlighting the differences between pre and post-commit versions. These error-modification pairs, detailed in Table~\ref{tab:error_c}, were crucial for developing prompts for LLM and for terms used in pattern matching, showcasing common human-made errors such as misuse of assignments and the mismatch port in instantiation. 
This approach showcases common human-made errors, such as misuse of assignments and mismatches in port instantiation, and provides a crucial benchmark for evaluating the verification effectiveness.

\begin{table}
    % \sffamily
    % \small
    \large
    \centering
    \vspace{-5pt}
     \caption{Part of common Verilog errors in real-world designs.} %\textcolor{red}{check the  descriptions and examples as I don't have domain knowledge}}
     % \vspace{-2pt}
     \label{Table:VerilogErrorsV2}
    \resizebox{1.0\linewidth}{!}{%
    \begin{tabular}{p{0.6cm} p{3.4cm}<{\centering} p{6.9cm}<{\centering}} 
    \toprule
    \rowcolor{white} \textbf{Types} & \textbf{Error} & \textbf{Symptoms} \\ 
    \midrule
    \multirow{4}{*}{\rotatebox[origin=c]{90}{\parbox[c]{15mm}{\centering \textbf{\\Declare}}}}& \multirow{2}{*}{\makecell[l]{Type Misuse}} & \cellcolor{green!15}\makecell[l]{output \textcolor{darkgreen}{\textbf{reg}} [8:0] result;}\\
    &&\cellcolor{red!15} \makecell[l]{output \textcolor{red}{\textbf{\quad \ \ }}[8:0] result;}\\ \cmidrule{2-3}
    & \multirow{2}{*}{\makecell[l]{Bitwidth Misuse} }&  \cellcolor{green!15}\makecell[l]{reg \textcolor{darkgreen}{\textbf{[8:0]}} count;} \\
    &&\cellcolor{red!15}\makecell[l]{reg \textcolor{red}{\textbf{[7:0]}} count; } \\ 
    \midrule
    \multirow{6}{*}{\rotatebox[origin=c]{90}{\parbox[c]{27mm}{\centering \textbf{\\Assignment}}}}& \multirow{2}{*}{\makecell[l]{Operator Misuse}} & \cellcolor{green!15}\makecell[l]{always @(*) result = a \textcolor{darkgreen}{\textbf{+}} b;}\\
    &&\cellcolor{red!15}\makecell[l]{always @(*) result = a \textcolor{red}{\textbf{-}} b;} \\ \cmidrule{2-3} 
    &\multirow{2}{*}{\makecell[l]{Variable Name Misuse}} &\cellcolor{green!15} \makecell[l]{assign r1 = \textcolor{darkgreen}{\textbf{r1\_temp}};}\\
    &&\cellcolor{red!15}\makecell[l]{assign r1 = \textcolor{red}{\textbf{r2\_temp}};} \\ \cmidrule{2-3} 
    & \multirow{2}{*}{\makecell[l]{Value Misuse}} & \cellcolor{green!15}\makecell[l]{if (rstn) data = \textcolor{darkgreen}{\textbf{32'b0}};}\\
    &&\cellcolor{red!15}\makecell[l]{if (rstn) data = \textcolor{red}{\textbf{32'b1}};} \\ 
    \midrule
        \multirow{4}{*}{\rotatebox[origin=c]{90}{\parbox[c]{15mm}{\centering \textbf{\\Condition}}}}& \multirow{2}{*}{\makecell[l]{Wrong Judgment Value}}& \cellcolor{green!15}\makecell[l]{for(i = 0; \textcolor{darkgreen}{\textbf{i $<$ 7}}; i ++) begin ... end}\\
        &&\cellcolor{red!15}\makecell[l]{for(i = 0; \textcolor{red}{\textbf{i $<$ 15}}; i ++) begin ... end}\\ \cmidrule{2-3}
    &\multirow{2}{*}{\makecell[l]{Wrong Sensitivity}} & \cellcolor{green!15}\makecell[l]{always@(posedge clk \textcolor{darkgreen}{\textbf{or negedge rstn}}) ... } \\
    &&\cellcolor{red!15}\makecell[l]{always@(posedge clk ) ... } \\ 
    \midrule
    \multirow{2}{*}{\rotatebox[origin=c]{90}{\parbox[c]{5mm}{\centering \textbf{\\Port}}}}& \multirow{2}{*}{\makecell[l]{Port Mismatch}}& \cellcolor{green!15}\makecell[l]{mod mod1(.a(a), .b(b), .in\_bd(\textcolor{darkgreen}{\textbf{\{bdg, 1'b1\}}}));}\\
    &&\cellcolor{red!15}\makecell[l]{mod mod1(.a(a), .b(b), .in\_bd(\textcolor{red}{\textbf{1'b1}}));}\\ 
        \bottomrule
    \end{tabular}
    } 
    % \vspace{-10pt}
    \label{tab:error_c}
\end{table}

\vspace{-2.5pt}
\section{Evaluation}
\label{sc:Evaluation}
\vspace{-2.5pt}
This section presents our experimental setup, evaluation metrics research questions, and discussions to the results. 

\parlabel{Setup.} 
In our experiment, we employed LLM agents via the OpenAI API, with GPT-4-turbo as the default model. 
An evaluation benchmark was then constructed using the extensively verified RTLLM dataset~\cite{lu2023rtllm}, which encompasses a diverse array of real-world errors. The initial code's ability to pass the compiler is indicated by a syntax error or functional error. 
% The benchmark consists of 10 common design types: Accumulator (``accu''); Adder (``adder\_8bit'', ``adder\_16bit'', ``adder\_pipe\_64bit''); Divider (``div\_16bit'', ``radix2\_div''); Multiplier (``multi\_booth\_8bit'', ``multi\_pipe\_8bit'', ``multi\_16bit''); Counter (``counter\_12'', ``JC\_counter''); FSM (``fsm'', ``pulse\_detect''); Memory (``asyn\_fifo'', ``right\_shifter''); Signal processor (``edge\_detect'', ``freq\_div'', ``parallel2serial'', ``serial2parallel'', ``signal\_generator'', ``synchronizer'', ``width\_8to16''); RISC-V (``alu'', ``pe'', ``RAM\_wr''); Schedulers (``calendar'', ``traffic\_light'').
We utilized a range of simulation tools including VCS~\cite{vcs}, Iverilog~\cite{iverilog}, Modelsim~\cite{modelsim}, Yosys~\cite{yosys}, and the linting tool Verilator~\cite{verilator} to ensure comprehensive verification and analysis.
We set the threshold of iterations to 5, as the improvement is hardly observed after that. 
All experiments were conducted on an AMD EPYC 7763 2.45GHz CPU.
For each instance, we asked LLMs for 5 times to reduce the randomness of the response.

\begin{figure*}[t]
\hspace{-2pt}
\vspace{-5pt}
\begin{minipage}[t]{0.49\textwidth}
\centering
\includegraphics[width=0.95\linewidth]{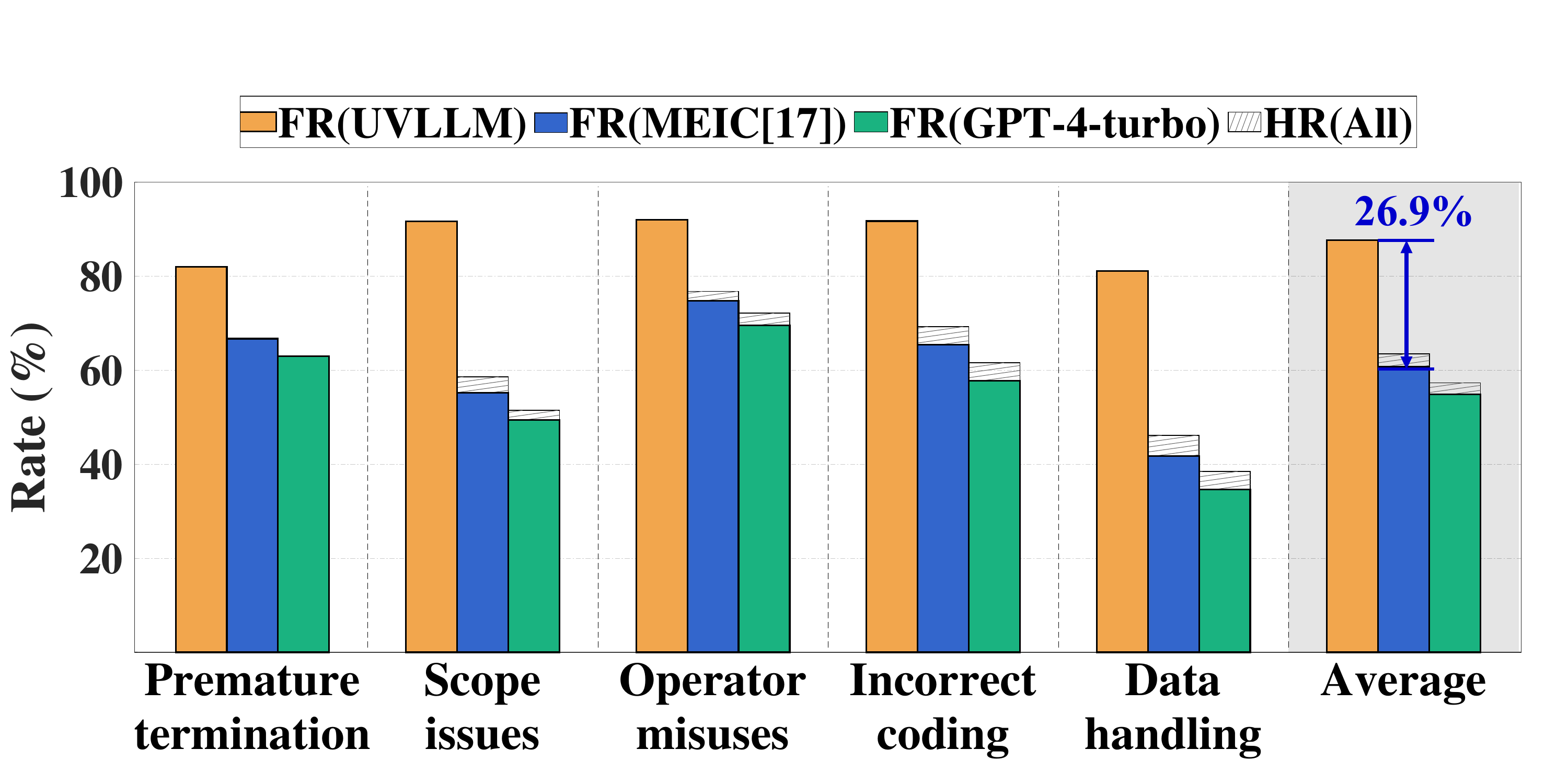}
\vspace{-3pt}
\caption{HR vs. FR in Syntax-Error Verification with Different Methods~\cite{xu2024meic}. The differences between HR and FR are shaded.}
\label{fig:res_syn}
\end{minipage}%
\hspace{5pt}
\begin{minipage}[t]{0.49\textwidth}
\centering
\includegraphics[width=0.95\linewidth]{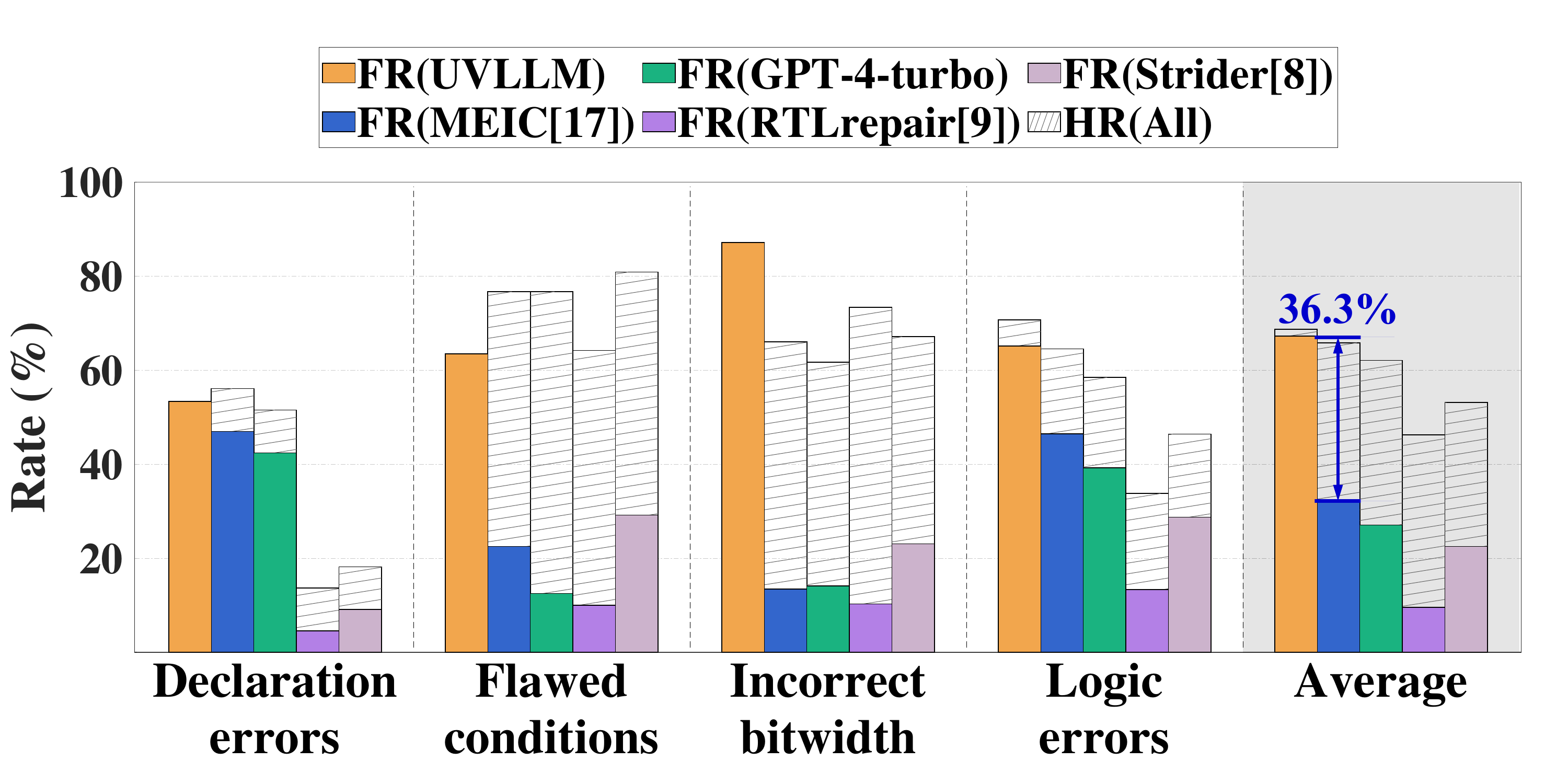}
\vspace{-3pt}
\caption{HR vs. FR in Functional-Error Verification with Different Methods~\cite{yang2024strider,laeufer2024rtl,xu2024meic}. The differences between HR and FR are shaded.}
\label{fig:res_func}
\end{minipage}
\vspace{-3pt}
\end{figure*}

\begin{figure}[t]
    \centering    
    \hspace{-15pt}
    \includegraphics[width=0.95\linewidth]{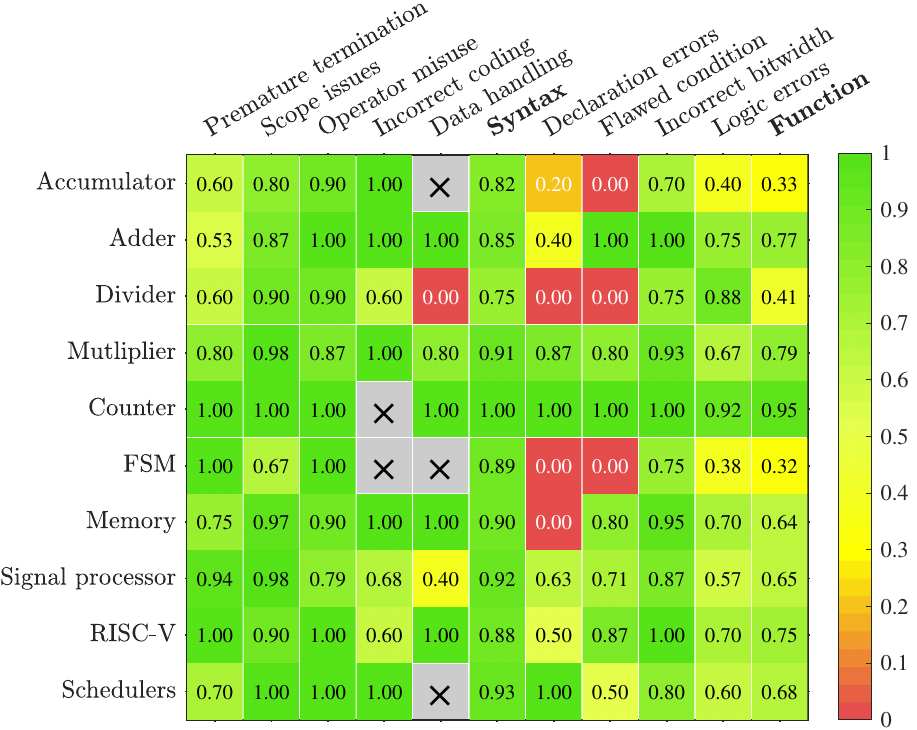}
    \caption{Heat map result for FR. The symbol ``×'' represents an error that could not be imposed due to the limitations of the specific module structure. Syntax and Function represent the weighted mean of the FR for syntax errors and function errors, respectively.}
    \label{fig:rlt1}
    \vspace{-2pt}
\end{figure}

\vspace{-1.5pt}
\subsection{Evaluation Metrics}%\textcolor{red}{implementation} Setup and 
% \vspace{-4pt}
\label{sbsc:EM}

 Recent work~\cite{chen2021evaluating,fakhoury2024llmbased}, tended to use pass@k metrics to assess functional correctness. 
For each problem in the problem set, k code samples are generated at a time, and the problem is considered solved if any sample passes the simulation test. 

% Specifically, we used Hit Rate (HR) to quantify the debug ability of the debugging framework\cite{tian2024debugbench}. 
% For an error code $\theta_{i}$ and its fixed version $\theta_{i}^{*}$, we had a corresponding set of test cases in testbench $\left(x_{i}^{0}, y_{i}^{0}\right)$,$\left(x_{i}^{1}, y_{i}^{1}\right)$, \ldots,$\left(x_{i}^{m}, y_{i}^{m}\right) $. 
% For the correct version of the code, $\theta_{i}^{*}$, it should produce the correct output $y_{i}^{j}$ when applied to the input data $x_{i}^{j}$ from the test cases. 
% That is, $a_{\theta_{i}^{*}}\left(x_{i}^{j}\right)=y_{i}^{j}$, the test case $\left(x_{i}^{j}, y_{i}^{j}\right) $ can be regarded as passing. 
% Whether the error is successfully fixed can be described as $\bigwedge_{j=0}^{m}\left[a_{\theta_{i}^{*}}\left(x_{i}^{j}\right)=y_{i}^{j}\right]$, an aggregate result of all test cases. 
% The HR that represents the test result on the bug instances are defined as:

% \begin{equation}
% \label{eq:hitrate}
% \text { \textbf{HR} }=\sum_{i=0}^{n} \frac{\bigwedge_{j=0}^{m}\left[a_{\theta_{i}^{*}}\left(x_{i}^{j}\right)=y_{i}^{j}\right]}{n} \times 100 \%
% \end{equation}

\parlabel{Hit Rate (HR).} Specifically, our framework quantifies effectiveness using Hit Rate (HR)~\cite{tian2024debugbench}. For erroneous code \( \theta_i \) and its corrected version \( \theta_i^* \), we evaluate a set of test cases \( \{(x_i^1, y_i^1), \ldots, (x_i^m, y_i^m)\} \). The corrected code \( \theta_i^* \) must produce the correct output \( y_i^j \) for each input \( x_i^j \), ensuring all cases pass. That is, $\bigwedge_{j=1}^{m}a_{\theta_{i}^{*}}\left(x_{i}^{j}\right)=y_{i}^{j}$. The overall rate for n corrcted versions is calculated as: 
\begin{equation}
\label{eq:hitrate}
\text { \textbf{HR} }=\sum_{i=1}^{n} \frac{\bigwedge_{j=1}^{m}\left[a_{\theta_{i}^{*}}\left(x_{i}^{j}\right)=y_{i}^{j}\right]}{n} \times 100 \%
\end{equation}
HR measures the proportion of instances resolved under test cases.

\parlabel{Fix Rate (FR).} 
% Nevertheless, given that the majority of contemporary testing methodologies rely on test cases, it can be reasonably deduced that fixes designated as "hit" will only fulfill a subset of the specified function points in the absence of comprehensive coverage.
% To address potential limitations in test coverage, our debugging framework incorporates Fix Rate (FR), a measure of expert validation~\cite{yang2024strider}. After a fix $\theta_{i}^{*}$ is proposed, experienced engineers review its correctness and applicability across a broader range of scenarios not covered by the initial test cases. This expert review culminates in the designation of $\hat{\theta}_{i}^{*}$ if the fix is confirmed, thereby quantifying the actual effectiveness of the debugging framework. 
% The FR that represents the actual repair of the bug instances are defined as:
To overcome limitations in test coverage, our framework includes Fix Rate (FR), which involves independent expert validation of the proposed fixes \( \theta_i^* \). After expert review, if the fix is confirmed effective across additional scenarios, it is designated as \( \hat{\theta}_i^* \):
\begin{equation}
\label{eq:fixrate}
\text { \textbf{FR} }=\sum_{i=1}^{n} \frac{\hat{\theta}_{i}^{*}}{n} \times 100 \%
\end{equation}
FR reflects the framework's repair effectiveness in broader conditions.

% It is worth noting that all HR and FR presented in this paper are calculated based on the average of 10 repeated experiments. 

\parlabel{Execution Time.} 
In evaluating the framework's performance, this paper emphasizes execution time as a critical metric, defined as the time interval from the initial design input into \name\ to the completion of the final code output. 

\vspace{-1.5pt}
\subsection{Results and Discussions}
\label{sbsc:eva}
\vspace{-1.5pt}

\noindent{\textbf{Result 1}}:The \approach\ framework’s effectiveness in enhancing hardware design verification is demonstrated through a comparative evaluation of Fix Rates (FRs) among various methodologies, as depicted in Figures \ref{fig:res_syn} and \ref{fig:res_func}. This analysis encompasses traditional script-based approaches~\cite{yang2024strider, laeufer2024rtl}, current LLM-aided methods~\cite{xu2024meic}, and a baseline incorporating GPT-4-turbo for benchmark repairs. Across all categories of errors—both syntax and functional—the \approach\ framework consistently achieves higher FRs.

Specifically, for syntax errors, the \approach\ framework records an average FR of 87.6\%, representing a significant 26.9\% improvement over the second performing method, MEIC. For functional errors, \approach\ continues to surpass other methods, registering a FR of 67.3\%, which is a 36.3\% enhancement relative to MEIC. Notably, in the case of \emph{Incorrect bitwidth} as illustrated in Fig.~\ref{fig:res_func}, \approach’s FR is more than double that of the next best method, RTLrepair. Moreover, even in less favorable scenarios, such as \emph{Declaration errors}, \approach’s FR remains approximately 5\% above that of MEIC, its nearest competitor. These results clearly demonstrate that \approach\ consistently outperforms other methods, achieving significantly higher FRs across diverse scenarios.

\noindent{\textbf{Result 2:}} The evaluation of the \approach\ framework’s capability to ensure that repaired code adheres to specification requirements is conducted through an examination of the correctness of repaired code. Although many methods achieve high hit rates (HRs), their fixes often overfit to specific input-output (IO) pairs, revealing a discrepancy between HRs and FRs.

Figures \ref{fig:res_syn} and \ref{fig:res_func} illustrate this disparity; the shaded areas represent the deviation between HR and FR, highlighting the failure of these methods to detect numerous errors, which leads to false negatives and insufficient coverage. Specifically, for syntax errors, \approach\ demonstrated no deviation across all scenarios, while deviations for other methods were observed in 4 out of 5 scenarios with an average of 5\% variations, confirming the achievement of high coverage for syntax errors. In contrast, for functional errors, the deviations for other methods were notably higher (all above 30\%) whereas \approach\ maintained a minimal deviation of only 1.4\%. Notably, \approach\ had a maximum deviation of just 5.6\% for \emph{Logic errors}, while the other methods displayed deviations exceeding 40\% for \emph{Flawed conditions}.

These results suggest that while other methods struggle to achieve comprehensive coverage for functional errors, \approach\ effectively mitigates this limitation. These findings indicate that \name\ significantly enhances the practicality of hardware design verification by integrating formal verification processes, to meet the specification requirements to the greatest extent possible.

\noindent{\textbf{Result 3:}} The \approach\ framework’s repair capabilities across a diverse range of hardware modules were evaluated by analyzing the FRs of 27 common modules, each injected with nine distinct types of syntax and functional errors. These modules were categorized into ten representative types, such as \emph{adders}, \emph{counters}, and \emph{FSMs}, to establish a comprehensive benchmark for evaluating the framework’s generalization in various verification scenarios.

As illustrated in Fig.~\ref{fig:rlt1}, where the framework’s FRs were depicted using color coding, \name\ exhibited exceptional adaptability, achieving robust FRs in simpler modules like \emph{counters}. For instance, the FRs for syntax errors and functional errors in these modules reached 100\% and 95\%, respectively. In contrast, the FRs were lower in more complex modules, such as \emph{FSMs}, with FRs for syntax errors and functional errors at 89\% and 32\%, respectively. This indicates that repairing more complex designs remains challenging.

Across the same types of module, syntax errors consistently exhibited higher FRs than functional errors, reflecting \name's proficiency in addressing syntactic issues.
This advantage stems from the extensive training of the LLM on a substantial corpus of HDL code data, enhancing its syntactic understanding. 
Additionally, the compiler and linter contribute detailed localization information that aids in the repair of syntax errors.
% while making strides in functional error correction.

Overall, the framework achieved an FR of 86.99\% for syntax errors and 71.92\% for functional errors, representing its reliability across diverse modules and error scenarios.

\begin{table*}
    % \sffamily
    \small
    \belowrulesep=0pt
    \aboverulesep=0pt
    \centering
     \caption{
     Performance comparison of segmented approach across common modules with various error instances. 
     % Errors are categorized as syntax ("s") and function ("f"). Modules are grouped as Arithmetic (Accumulator, Adder, Divider, Multiplier), Control (Counter, FSM), Memory, and Miscellaneous (other modules). The repair operation of \approach\ is seperated into Preprocessing, Repair in Mismatch Signal (MS) Mode, Repair in Suspicious Line (SL) Mode.
     }
     % \vspace{-2pt}
     \label{tab:seg}
     \begin{threeparttable}
    \resizebox{0.95\linewidth}{!}{%
    \begin{tabular}{c| c c  c c  c c  c c| c c | c }
    %p{1.2cm} p{1.2cm}| p{1.2cm} p{1.2cm}| p{1.2cm} p{1.2cm}| p{1.2cm} p{1.2cm}}
    \toprule
    % \multirow{3}{*}{\centering\textbf{Types}} &\multicolumn{8}{c|}{\centering\textbf{\approach}}& \multicolumn{2}{c}{\multirow{1}{*}{\centering\textbf{MEIC\cite{xu2024meic}}}}&  \multirow{3}{*}{\centering\textbf{Speedup}}\\ %\cline{2-9}
    \multirow{2}{*}{\centering\textbf{Types}} & \multicolumn{2}{c}{\centering\textbf{Pre-processing\tnote{1}}} & \multicolumn{2}{c}{\centering\textbf{Repair in MS Mode}} & \multicolumn{2}{c|}{\centering\textbf{Repair in SL Mode}} & \multicolumn{2}{c|}{\centering\textbf{\approach \tnote{2}}} & \multicolumn{2}{c|}{\centering\textbf{MEIC~\cite{xu2024meic}}} & \multirow{2}{*}{\centering\textbf{Speedup}} \\
    &  $FR/\%$& $T_{exec}/s$ & $FR/\%$& $T_{exec}/s$ & $FR/\%$& \multicolumn{1}{c|}{$T_{exec}/s$} & $FR/\%$& $T_{exec}/s$ & $FR/\%$& $T_{exec}/s$&  \\
    \midrule

    Arithmetic\tnote{3}\ \ \  s\tnote{4}& 69.93  & \multicolumn{1}{c|}{\ \ 8.30 }& 13.07  & \multicolumn{1}{c|}{\ \ 5.60} & 1.31  & \multicolumn{1}{c|}{0.30}& 84.31{\cellcolor[HTML]{85e23d}} & 14.20 &62.30{\cellcolor[HTML]{bff73e}}  &197.29 &\textbf{13.89x}\\ 
    Control s& 80.91  & \multicolumn{1}{c|}{\ \  7.23}&\ \  8.18  & \multicolumn{1}{c|}{\ \ 3.38}& 0.00  & \multicolumn{1}{c|}{0.00}& 89.09{\cellcolor[HTML]{71E822}} & 10.61&63.64{\cellcolor[HTML]{c7f942}}  &129.02 &\textbf{12.61x}\\ 
    Memory s& 60.00  & \multicolumn{1}{c|}{10.29}& 28.33  & \multicolumn{1}{c|}{\ \ 5.14}& 0.00  & \multicolumn{1}{c|}{0.00}& 88.33{\cellcolor[HTML]{85e23d}} & 15.43 &54.55{\cellcolor[HTML]{dbfc49}}  &147.17 & \ \ 9.53x\\ 
    Miscellaneous s& 79.65  & \multicolumn{1}{c|}{\ \ 8.31}& \ \ 7.67  & \multicolumn{1}{c|}{\ \ 4.77}& 1.18  & \multicolumn{1}{c|}{0.39}& 88.50{\cellcolor[HTML]{85e23d}} & 13.47 &66.67{\cellcolor[HTML]{b5f53c}}  &\ \ 62.67 & \ \ 4.65x\\ 
    \textbf{Syntax}& \textbf{74.72} &\multicolumn{1}{c|}{\ \ 8.49} & 11.29 & \multicolumn{1}{c|}{\ \ 5.06}& 0.98 & \multicolumn{1}{c|}{0.27}& \textbf{86.99}{\cellcolor[HTML]{85e23d}} & \multicolumn{1}{c|}{\textbf{13.83}}&\textbf{62.99}{\cellcolor[HTML]{bff73e}}  &\textbf{134.95} &\ \ 9.76x\\ 
    \midrule
    
    Arithmetic f& 30.26  & \multicolumn{1}{c|}{\ \ 3.63}& 33.33  & \multicolumn{1}{c|}{11.27}& 2.63  & \multicolumn{1}{c|}{0.64}& 66.23{\cellcolor[HTML]{b5f53c}} & 15.54 &40.53{\cellcolor[HTML]{f4ff1f}}  &257.28 &\textbf{16.56x}\\ 
    Control f& 29.93  & \multicolumn{1}{c|}{\ \ 3.33}& 30.61  & \multicolumn{1}{c|}{\ \ 9.37}& 5.44  & \multicolumn{1}{c|}{0.84}& 65.99{\cellcolor[HTML]{b5f53c}} & 13.54&10.91{\cellcolor[HTML]{ee8833}}  &163.96 & \textbf{12.55x}\\ 
    Memory f& 25.00  & \multicolumn{1}{c|}{\ \ 4.35}& 58.33  & \multicolumn{1}{c|}{11.22}& 3.33  & \multicolumn{1}{c|}{0.87}& 86.67{\cellcolor[HTML]{85e23d}} & 16.44 &22.73{\cellcolor[HTML]{f6c318}}  &256.49 & \textbf{15.60x}\\ 
    Miscellaneous f& 21.25  & \multicolumn{1}{c|}{\ \ 3.86}& 49.06  & \multicolumn{1}{c|}{11.54}& 5.63  & \multicolumn{1}{c|}{0.90}& 75.94{\cellcolor[HTML]{99f030}} & 16.30 &40.07{\cellcolor[HTML]{f4ff1f}}  &\ \ 65.37 &\ \ 4.01x\\ 
    \textbf{Function}& 25.96 &\multicolumn{1}{c|}{\ \ 3.82} & \textbf{41.46} & \multicolumn{1}{c|}{11.19}& 4.50 & \multicolumn{1}{c|}{0.78}& \textbf{71.92}{\cellcolor[HTML]{99f030}} & \multicolumn{1}{c|}{\textbf{15.79}}&\textbf{34.57}{\cellcolor[HTML]{fbff0a}}  &\textbf{191.76} &\textbf{12.14x}\\
    \bottomrule
     \emph{Overall}& \textbf{51.27}  & \multicolumn{1}{c|}{\ \ \textbf{6.16}}& \textbf{25.80}  & \multicolumn{1}{c|}{\ \ \textbf{7.79}}& \textbf{2.68}  & \multicolumn{1}{c|}{\textbf{0.49}}& \textbf{79.75}{\cellcolor[HTML]{8fef2d}} & \textbf{14.77} &\textbf{52.14}{\cellcolor[HTML]{dbfc49}}  &\textbf{153.84} &\textbf{10.42x}\\
    \bottomrule
    \end{tabular}
    }
    \begin{tablenotes}
        \footnotesize
        \item[1] The repair operation of \approach\ comprises three stages: Pre-processing, Repair in Mismatch Signal (MS) Mode, and Repair in Suspicious Line (SL) Mode. The columns labeled $FR$ and $T_{exec}$ indicate the contributions of each stage to the fix rate and execution time, respectively. 
        \item[2] The column labeled \name\ summarizes the total contributions across all stages of the repair operation.
        \item[3] Modules are grouped as Arithmetic (Accumulator, Adder, Divider, Multiplier), Control (Counter, FSM), Memory, and Miscellaneous (other modules).
        \item[4] Errors are categorized as syntax (``s'') and function (``f'').  
    \end{tablenotes}
    \end{threeparttable}
\vspace{-15pt}
\end{table*}

\noindent{\textbf{Result 4:}} The \name\ framework’s evaluation primarily focuses on how its distinct stages contribute to the fix rate and execution time during the repair operation, providing insights into each segment of the verification process.

Table~\ref{tab:seg} details the repair process, which unfolds in several stages, each playing a different role in resolving syntax and functional errors. The Pre-processing stage was particularly effective in addressing syntax errors, successfully resolving 74.72\% of these cases as highlighted. For functional errors, the Mismatch Signal (MS) mode in the Repair stage was most effective, correcting 41.46\% of the instances. The adoption of segmented steps enables \approach\ to work effectively and adapt flexibly to various verification scenarios.

% \noindent \textbf{(Segmentability). How do different stages of \name\ framework contribute to the overall process?} 
% How does \name\ perform at different stages of the repair process?
% This research question mainly investigates the contributions of distinct stages to fix rate and execution time achieved by \name\ during the repair operation, providing insights into the segmentability of the entire verification process. 

% As detailed in Table~\ref{tab:seg}, the repair process unfolds in several stages, each contributing differently to the resolution of syntax and functional errors. The Pre-processing stage demonstrated remarkable contributions to address syntax errors, successfully resolving 74.72\% of instances. For functional errors, the Mismatch Signal mode in the Repair stage dominated, rectifying 41.46\% of the cases.
% The adoption of segmented steps enables \approach\ to work effectively and adapt flexibly to various verification scenarios.

For errors strictly related to syntax, the majority were successfully corrected during the pre-processing stage. However, 11.29\% of syntax-only errors persisted and advanced to the subsequent repair stage in MS mode. Similarly, the attempt to resolve 25.96\% of functional errors inadvertently introduced new syntax issues, which were then addressed by the pre-processor. This demonstrates \approach's ability to compensate for new errors introduced in earlier stages and mitigate the uncertainties associated with LLMs.
% When considering the execution time of the segmented repair operation, it can be observed though pre-processing accounts for repairing more than 50\% of the overall benchmark, it executes less time than repair in MS mode, highlighting the effectiveness of introducing pre-processing stage.

In terms of execution time, the segmented repair operation shows that the pre-processing stage, despite handling over 50\% of all benchmark repairs, typically requires less time than repairs conducted in MS mode. This demonstrates the efficiency benefits of incorporating a robust pre-processing stage.

% When evaluating the execution time of the segmented repair operation, it is evident that the pre-processing stage, despite accounting for more than 50\% of all benchmark repairs, requires less time in average than the repairs conducted in MS mode. This underscores the efficiency gained by integrating  the pre-processing stage.
%As detailed in Table~\ref{Table:seg}, the distinct stages of the process exert differential impacts across various instances. Notably, the majority of syntax error instances are effectively addressed during the Preprocessing stage, while the Mismatch Signal (MS) mode predominantly rectifies the majority of functional errors in the Repair stage, accounting for 74.72\% and 41.46\% of the respective cases. For errors solely related to syntax, most instances are successfully corrected in the preprocessing phase. However, there are instances where syntax-only errors persist and progress through subsequent stages of the process. Similarly, the process of resolving functional errors may inadvertently lead to the emergence of new syntax issues. This occurrence can primarily be attributed to the innovative capabilities of LLMs, which, while attempting to optimize the modules, may inadvertently introduce new errors.

% \noindent \textbf{(Performability). How does the \name\ framework compare with existing methods in execution performance?}
% This research question focuses on evaluating the framework’s execution efficiency compared to existing methods from a multidimensional perspective. 

\noindent{\textbf{Result 5:}} The \name\ framework’s execution efficiency was evaluated against existing methods from a multidimensional perspective, mainly focusing on two key metrics: Failure Rates (FRs) and execution time $T_{exec}$, as detailed in Table~\ref{tab:seg}.

While Fig.\ref{fig:res_syn} and Fig.\ref{fig:res_func} show the variations in FR performance across different error types, \name\ consistently surpassed MEIC across all module types. For instance, within the \emph{Miscellaneous} modules for syntax errors in Table~\ref{tab:seg}, \name\ achieved an FR of 88.50\%, which is 21.83\% higher than MEIC’s 66.67\%. In handling functional errors within \emph{Arithmetic} modules, characterized by their complex logic, \name\ maintained an FR of 66.23\%, significantly outperforming MEIC’s 40.53\%. These results underscore \name’s robust capability to effectively resolve a wide spectrum of errors across different modules compared to existing methods, thus proving its effectiveness and reliability in boosting system performance.

In terms of operational efficiency, \name\ also demonstrated a substantial reduction in execution time. For example, when processing syntax errors within the \emph{Miscellaneous} modules, \name\ recorded an average execution time of 13.47s, marking a speedup of 4.65x compared with MEIC. This advantage was even more pronounced when addressing complex functional errors, where \name\ achieved a speedup of up to 16.56x over MEIC in \emph{Arithmetic} modules. On average, \name\ operated 10.42x faster than MEIC, while simultaneously achieving higher test coverage and pass rates. These findings highlight \name’s potential to significantly enhance the design verification process for practical deployment by merging increased automation with superior efficiency.

\begin{table}[t]
    % \sffamily
    \small
    \belowrulesep=0pt
    \aboverulesep=0pt
    \centering
        \caption{Ablation study. UVLLM$_{pair}$ and UVLLM$_{comp}$ represent UVLLM with LLMs generating code pairs and complete codes.}
    \resizebox{0.8\linewidth}{!}{%
    \begin{tabular}{c|c c|c c}
    \toprule
    \multirow{2}{*}{\centering\textbf{Framework}} & \multicolumn{2}{c|}{\centering\textbf{$FR/\%$}} & \multicolumn{2}{c}{\centering\textbf{$T_{exec}/s$}} \\
    &  Syntax& Func.& Syntax& Func.\\
    % Framework&$FR_{syntax}$&$T_{exec,syntax}$&$FR_{func}$&$T_{exec,func}$\\
    \midrule
         $UVLLM_{pair}$& 86.99 & 71.92 & 13.83 & 15.79\\
         $UVLLM_{comp}$& 70.41 & 59.25 & 35.60 & 71.84\\
         \bottomrule
    \end{tabular}}
    % \vspace{-3pt}
    \label{tab:abla1}
\end{table}

\vspace{-3.5pt}
\subsection{Ablation Study}
Our research includes the ablation study designed to evaluate the impact of iteration strategies on the effectiveness of the framework.

\parlabel{Repair generation form.} We initially employ an approach that uses original-repair code pairs to facilitate the generation of new code by leveraging outputs from LLMs. However, the ablation study examines an alternative method where entire code snippets are directly produced by the LLMs, omitting the generation of repair pairs.

As shown in Table~\ref{tab:abla1}, generating complete code snippets resulted in a slight decline in repair accuracy and an increase in execution time compared to generating original-repair code pairs. Nevertheless, there were specific scenarios where this direct generation method proved superior. This advantage is mainly due to the ability of the direct generation method to handle minor errors that pose significant challenges for LLMs in terms of search efficiency. In some cases, regenerating the entire code is more effective than trying to modify or replace segments of the existing code. For instance, correcting the error ``$module\ a(A);... (Missing\ Definition\ of \ Port\ A) ... endmodule$'' proves challenging for the replacement strategy, primarily because the essential context is frequently overlooked, whereas the reproduction method handles it more straightforwardly.

\vspace{-1.5pt}
\section{Conclusion}
\vspace{-1.5pt}
\label{sc:Conclusion}
In this work, an automated universal verification framework, \textbf{\approach}, which comprehensively addresses main phases of hardware design verification, including testbench construction, test execution, result analysis, and repair, is proposed. 
% By fulfill the complete phases in the hardware verification (\ie, testbench construction, test execution, result analysis and repair), 
Tested on the proposed benchmark, \approach\ achieves average syntax and functional error fault rates of 86.99\% and 71.92\%, respectively, while maintaining nearly 100\% test coverage. Additionally, \approach\ performs 10.42 times faster than the previous MEIC framework.
The framework demonstrates that it is feasible to employ the LLMs for the purpose of Verilog code verification, irrespective of the initial code state. The utilization of reasonable segmentation and feedback engineering leads to an improvement in the verification efficiency of the framework. 
\clearpage
\bibliographystyle{IEEEtran}
\bibliography{ref}

\begin{thebibliography}{10}
\providecommand{\url}[1]{#1}
\csname url@samestyle\endcsname
\providecommand{\newblock}{\relax}
\providecommand{\bibinfo}[2]{#2}
\providecommand{\BIBentrySTDinterwordspacing}{\spaceskip=0pt\relax}
\providecommand{\BIBentryALTinterwordstretchfactor}{4}
\providecommand{\BIBentryALTinterwordspacing}{\spaceskip=\fontdimen2\font plus
\BIBentryALTinterwordstretchfactor\fontdimen3\font minus \fontdimen4\font\relax}
\providecommand{\BIBforeignlanguage}[2]{{%
\expandafter\ifx\csname l@#1\endcsname\relax
\typeout{** WARNING: IEEEtran.bst: No hyphenation pattern has been}%
\typeout{** loaded for the language `#1'. Using the pattern for}%
\typeout{** the default language instead.}%
\else
\language=\csname l@#1\endcsname
\fi
#2}}
\providecommand{\BIBdecl}{\relax}
\BIBdecl

\bibitem{lahti2018we}
S.~Lahti, P.~Sj{\"o}vall, J.~Vanne, and T.~D. H{\"a}m{\"a}l{\"a}inen, ``Are we there yet? a study on the state of high-level synthesis,'' \emph{IEEE Transactions on Computer-Aided Design of Integrated Circuits and Systems}, vol.~38, no.~5, pp. 898--911, 2018.

\bibitem{liu2018survey}
Y.~Liu, L.~Zhang, and Z.~Zhang, ``A survey of test based automatic program repair.'' \emph{J. Softw.}, vol.~13, no.~8, pp. 437--452, 2018.

\bibitem{liu2021critical}
K.~Liu, L.~Li, A.~Koyuncu, D.~Kim, Z.~Liu, J.~Klein, and T.~F. Bissyand{\'e}, ``A critical review on the evaluation of automated program repair systems,'' \emph{Journal of Systems and Software}, vol. 171, p. 110817, 2021.

\bibitem{zhang2023critical}
Q.~Zhang, T.~Zhang, J.~Zhai, C.~Fang, B.~Yu, W.~Sun, and Z.~Chen, ``A critical review of large language model on software engineering: An example from chatgpt and automated program repair,'' \emph{arXiv preprint arXiv:2310.08879}, 2023.

\bibitem{yin2024thinkrepair}
X.~Yin, C.~Ni, S.~Wang, Z.~Li, L.~Zeng, and X.~Yang, ``Thinkrepair: Self-directed automated program repair,'' in \emph{Proceedings of the 33rd ACM SIGSOFT International Symposium on Software Testing and Analysis}, 2024, pp. 1274--1286.

\bibitem{xu2024automated}
K.~Xu, G.~L. Zhang, X.~Yin, C.~Zhuo, U.~Schlichtmann, and B.~Li, ``Automated c/c++ program repair for high-level synthesis via large language models,'' in \emph{Proceedings of the 2024 ACM/IEEE International Symposium on Machine Learning for CAD}, 2024, pp. 1--9.

\bibitem{ahmad2022cirfix}
H.~Ahmad, Y.~Huang, and W.~Weimer, ``Cirfix: automatically repairing defects in hardware design code,'' in \emph{Proceedings of the 27th ACM International Conference on Architectural Support for Programming Languages and Operating Systems}, 2022, pp. 990--1003.

\bibitem{yang2024strider}
D.~Yang, J.~He, X.~Mao, T.~Li, Y.~Lei, X.~Yi, and J.~Wu, ``Strider: Signal value transition-guided defect repair for hdl programming assignments,'' \emph{IEEE Transactions on Computer-Aided Design of Integrated Circuits and Systems}, vol.~43, no.~5, pp. 1594--1607, 2024.

\bibitem{laeufer2024rtl}
K.~Laeufer, B.~Fajardo, A.~Ahuja, V.~Iyer, B.~Nikoli{\'c}, and K.~Sen, ``Rtl-repair: Fast symbolic repair of hardware design code,'' in \emph{Proceedings of the 29th ACM International Conference on Architectural Support for Programming Languages and Operating Systems, Volume 3}, 2024, pp. 867--881.

\bibitem{liu2023verilogeval}
M.~Liu, N.~Pinckney, B.~Khailany, and H.~Ren, ``Verilogeval: Evaluating large language models for verilog code generation,'' in \emph{2023 IEEE/ACM International Conference on Computer Aided Design (ICCAD)}.\hskip 1em plus 0.5em minus 0.4em\relax IEEE, 2023, pp. 1--8.

\bibitem{thakur2023verigen}
S.~Thakur, B.~Ahmad, H.~Pearce, B.~Tan, B.~Dolan-Gavitt, R.~Karri, and S.~Garg, ``Verigen: A large language model for verilog code generation,'' \emph{arXiv preprint arXiv:2308.00708}, 2023.

\bibitem{blocklove2023chip}
J.~Blocklove, S.~Garg, R.~Karri, and H.~Pearce, ``Chip-chat: Challenges and opportunities in conversational hardware design,'' \emph{arXiv preprint arXiv:2305.13243}, 2023.

\bibitem{delorenzo2024make}
M.~DeLorenzo, A.~B. Chowdhury, V.~Gohil, S.~Thakur, R.~Karri, S.~Garg, and J.~Rajendran, ``Make every move count: Llm-based high-quality rtl code generation using mcts,'' \emph{arXiv preprint arXiv:2402.03289}, 2024.

\bibitem{liu2023chipnemo}
M.~Liu, T.-D. Ene, R.~Kirby, C.~Cheng, N.~Pinckney, R.~Liang, J.~Alben, H.~Anand, S.~Banerjee, I.~Bayraktaroglu \emph{et~al.}, ``Chipnemo: Domain-adapted llms for chip design,'' \emph{arXiv preprint arXiv:2311.00176}, 2023.

\bibitem{tsai2023rtlfixer}
Y.~Tsai, M.~Liu, and H.~Ren, ``Rtlfixer: Automatically fixing rtl syntax errors with large language models,'' \emph{arXiv preprint arXiv:2311.16543}, 2023.

\bibitem{yao2024hdldebugger}
X.~Yao, H.~Li, T.~H. Chan, W.~Xiao, M.~Yuan, Y.~Huang, L.~Chen, and B.~Yu, ``Hdldebugger: Streamlining hdl debugging with large language models,'' \emph{arXiv preprint arXiv:2403.11671}, 2024.

\bibitem{xu2024meic}
K.~Xu, J.~Sun, Y.~Hu, X.~Fang, W.~Shan, X.~Wang, and Z.~Jiang, ``Meic: Re-thinking rtl debug automation using llms,'' \emph{arXiv preprint arXiv:2405.06840}, 2024.

\bibitem{fu2023llm4sechw}
W.~Fu, K.~Yang, R.~G. Dutta, X.~Guo, and G.~Qu, ``Llm4sechw: Leveraging domain-specific large language model for hardware debugging,'' in \emph{2023 Asian Hardware Oriented Security and Trust Symposium (AsianHOST)}.\hskip 1em plus 0.5em minus 0.4em\relax IEEE, 2023, pp. 1--6.

\bibitem{white2023prompt}
J.~White, Q.~Fu, S.~Hays, M.~Sandborn, C.~Olea, H.~Gilbert, A.~Elnashar, J.~Spencer-Smith, and D.~C. Schmidt, ``A prompt pattern catalog to enhance prompt engineering with chatgpt,'' \emph{arXiv preprint arXiv:2302.11382}, 2023.

\bibitem{sahoo2024systematic}
P.~Sahoo, A.~K. Singh, S.~Saha, V.~Jain, S.~Mondal, and A.~Chadha, ``A systematic survey of prompt engineering in large language models: Techniques and applications,'' \emph{arXiv preprint arXiv:2402.07927}, 2024.

\bibitem{wei2022chain}
J.~Wei, X.~Wang, D.~Schuurmans, M.~Bosma, F.~Xia, E.~Chi, Q.~V. Le, D.~Zhou \emph{et~al.}, ``Chain-of-thought prompting elicits reasoning in large language models,'' \emph{Advances in neural information processing systems}, vol.~35, pp. 24\,824--24\,837, 2022.

\bibitem{openai2024gptapi}
\BIBentryALTinterwordspacing
OpenAI, ``{GPT-4 pricing for OpenAI API},'' 2024. [Online]. Available: \url{https://openai.com/api/pricing/}
\BIBentrySTDinterwordspacing

\bibitem{plasencia2018robust}
F.~Plasencia-Balabarca, E.~Mitacc-Meza, M.~Raffo-Jara, and C.~Silva-C{\'a}rdenas, ``Robust functional verification framework based in uvm applied to an aes encryption module,'' in \emph{2018 New Generation of CAS (NGCAS)}.\hskip 1em plus 0.5em minus 0.4em\relax IEEE, 2018, pp. 194--197.

\bibitem{spear2008systemverilog}
C.~Spear, \emph{SystemVerilog for Verification: A Guide to Learning the Testbench Language Features}.\hskip 1em plus 0.5em minus 0.4em\relax Springer, 2008, vol. 161.

\bibitem{semeria2000methodology}
L.~S{\'e}m{\'e}ria and A.~Ghosh, ``Methodology for hardware/software co-verification in c/c++ (short paper),'' in \emph{Proceedings of the 2000 Asia and South Pacific Design Automation Conference}, 2000, pp. 405--408.

\bibitem{nakamura2004fast}
Y.~Nakamura, K.~Hosokawa, I.~Kuroda, K.~Yoshikawa, and T.~Yoshimura, ``A fast hardware/software co-verification method for system-on-a-chip by using a c/c++ simulator and fpga emulator with shared register communication,'' in \emph{Proceedings of the 41st annual Design Automation Conference}, 2004, pp. 299--304.

\bibitem{radu2024generative}
V.~Radu, D.~Dranga, C.~Dumitrescu, A.~I. Tabirca, and M.~C. Stefan, ``Generative ai assertions in uvm-based system verilog functional verification,'' \emph{Systems}, vol.~12, no.~10, p. 390, 2024.

\bibitem{ji2023towards}
Z.~Ji, T.~Yu, Y.~Xu, N.~Lee, E.~Ishii, and P.~Fung, ``Towards mitigating llm hallucination via self reflection,'' in \emph{Findings of the Association for Computational Linguistics: EMNLP 2023}, 2023, pp. 1827--1843.

\bibitem{amatriain2024measuring}
X.~Amatriain, ``Measuring and mitigating hallucinations in large language models: amultifaceted approach,'' 2024.

\bibitem{galitsky2023truth}
B.~A. Galitsky, ``Truth-o-meter: Collaborating with llm in fighting its hallucinations,'' 2023.

\bibitem{chen2023hallucination}
Y.~Chen, Q.~Fu, Y.~Yuan, Z.~Wen, G.~Fan, D.~Liu, D.~Zhang, Z.~Li, and Y.~Xiao, ``Hallucination detection: Robustly discerning reliable answers in large language models,'' in \emph{Proceedings of the 32nd ACM International Conference on Information and Knowledge Management}, 2023, pp. 245--255.

\bibitem{ji2023survey}
Z.~Ji, N.~Lee, R.~Frieske, T.~Yu, D.~Su, Y.~Xu, E.~Ishii, Y.~J. Bang, A.~Madotto, and P.~Fung, ``Survey of hallucination in natural language generation,'' \emph{ACM Computing Surveys}, vol.~55, no.~12, pp. 1--38, 2023.

\bibitem{lambert2024rewardbench}
N.~Lambert, V.~Pyatkin, J.~Morrison, L.~Miranda, B.~Y. Lin, K.~Chandu, N.~Dziri, S.~Kumar, T.~Zick, Y.~Choi \emph{et~al.}, ``Rewardbench: Evaluating reward models for language modeling,'' \emph{arXiv preprint arXiv:2403.13787}, 2024.

\bibitem{Structured-outputs}
\BIBentryALTinterwordspacing
OpenAI, ``{Structured outputs - OpenAI API},'' 2024. [Online]. Available: \url{https://platform.openai.com/docs/guides/structured-outputs}
\BIBentrySTDinterwordspacing

\bibitem{sudakrishnan2008understanding}
S.~Sudakrishnan, J.~Madhavan, E.~J. Whitehead~Jr, and J.~Renau, ``Understanding bug fix patterns in verilog,'' in \emph{Proceedings of the 2008 international working conference on Mining software repositories}, 2008, pp. 39--42.

\bibitem{ma2022debugging}
J.~Ma, G.~Zuo, K.~Loughlin, H.~Zhang, A.~Quinn, and B.~Kasikci, ``Debugging in the brave new world of reconfigurable hardware,'' in \emph{Proceedings of the 27th ACM International Conference on Architectural Support for Programming Languages and Operating Systems}, 2022, pp. 946--962.

\bibitem{antinyan2017evaluating}
V.~Antinyan, M.~Staron, and A.~Sandberg, ``Evaluating code complexity triggers, use of complexity measures and the influence of code complexity on maintenance time,'' \emph{Empirical Software Engineering}, vol.~22, pp. 3057--3087, 2017.

\bibitem{lu2023rtllm}
Y.~Lu, S.~Liu, Q.~Zhang, and Z.~Xie, ``Rtllm: An open-source benchmark for design rtl generation with large language model,'' \emph{arXiv preprint arXiv:2308.05345}, 2023.

\bibitem{vcs}
\BIBentryALTinterwordspacing
Synopsys, ``{VCS: Synopsys Verification Continuum},'' 2024. [Online]. Available: \url{https://www.synopsys.com/verification/simulation/vcs.html}
\BIBentrySTDinterwordspacing

\bibitem{iverilog}
\BIBentryALTinterwordspacing
I.~V. Team, ``{Icarus Verilog},'' 2024. [Online]. Available: \url{http://iverilog.icarus.com/}
\BIBentrySTDinterwordspacing

\bibitem{modelsim}
\BIBentryALTinterwordspacing
Siemens, ``{ModelSim},'' 2024. [Online]. Available: \url{https://eda.sw.siemens.com/en-US/ic/modelsim/}
\BIBentrySTDinterwordspacing

\bibitem{yosys}
\BIBentryALTinterwordspacing
C.~Wolf, ``{Yosys Open SYnthesis Suite},'' 2024. [Online]. Available: \url{http://www.clifford.at/yosys/}
\BIBentrySTDinterwordspacing

\bibitem{verilator}
\BIBentryALTinterwordspacing
W.~Snyder, ``{Verilator},'' 2024. [Online]. Available: \url{https://www.veripool.org/wiki/verilator}
\BIBentrySTDinterwordspacing

\bibitem{chen2021evaluating}
M.~Chen, J.~Tworek, H.~Jun, Q.~Yuan, H.~P. d.~O. Pinto, J.~Kaplan, H.~Edwards, Y.~Burda, N.~Joseph, G.~Brockman \emph{et~al.}, ``Evaluating large language models trained on code,'' \emph{arXiv preprint arXiv:2107.03374}, 2021.

\bibitem{fakhoury2024llmbased}
S.~Fakhoury, A.~Naik, G.~Sakkas, S.~Chakraborty, and S.~K. Lahiri, ``Llm-based test-driven interactive code generation: User study and empirical evaluation,'' 2024.

\bibitem{tian2024debugbench}
R.~Tian, Y.~Ye, Y.~Qin, X.~Cong, Y.~Lin, Z.~Liu, and M.~Sun, ``Debugbench: Evaluating debugging capability of large language models,'' \emph{arXiv preprint arXiv:2401.04621}, 2024.

\end{thebibliography}

\end{document}